\documentclass[aps,prl,twocolumn,groupedaddress]{revtex4}
\usepackage[T1]{fontenc}
\usepackage[latin9]{inputenc}
\usepackage{color}
\usepackage{amsmath}
\usepackage{amssymb}
\usepackage{graphicx}
\usepackage{braket}
\usepackage{pgfplots}
\usepackage{overpic}
\usepackage{subfigure}
\usepackage{capt-of}
\usepackage{pgfplots}
\pgfplotsset{width=7cm,compat=1.8}
\usepackage{lmodern}
\usepackage{soul}
\usepackage[nodayofweek,level]{datetime}

\renewcommand*{\vec}[1]{\mathbf{#1}}

\newcommand*{\im}{\, \mathrm{Im}}
\newcommand*{\re}{\, \mathrm{Re}}

\usetikzlibrary{positioning}
\usetikzlibrary{spy}

\makeatletter
\def\blfootnote{\xdef\@thefnmark{}\@footnotetext}

\usepackage{bbold}
\usepackage{tikz}

\usepackage{dsfont}
\RequirePackage[colorlinks=true,linkcolor=blue]{hyperref}

\makeatother

\usetikzlibrary{positioning}
\usetikzlibrary{calc}
\usetikzlibrary{arrows,shapes,backgrounds}

\usepackage{soul}

\begin{document}

\title{Wigner Crystals in Two-Dimensional Transition-Metal Dichalcogenides: Spin Physics and Readout}

\author{J. Kn\"orzer,$^{1,2,*}$ M. J. A. Schuetz,$^{3,\dagger}$ G. Giedke,$^{4,5}$ D. S. Wild,$^{3}$ K. De Greve,$^{3,\ddagger}$ R. Schmidt,$^{1,2}$ M. D. Lukin,$^{3}$ and J. I. Cirac$^{1,2}$}

\affiliation{$^{1}$Max-Planck-Institut f\"ur Quantenoptik, Hans-Kopfermann-Str. 1, D-85748 Garching, Germany}
\affiliation{$^{2}$Munich Center for Quantum Science and Technology (MCQST), Schellingstr. 4, D-80799 M\"unchen, Germany}
\affiliation{$^{3}$Physics Department, Harvard University, Cambridge, MA 02318, USA}
\affiliation{$^{4}$Donostia International Physics Center, Paseo Manuel de Lardizabal 4, E-20018 San Sebasti\'{a}n, Spain}
\affiliation{$^{5}$Ikerbasque Foundation for Science, Maria Diaz de Haro 3, E-48013 Bilbao, Spain}
\thanks{\noindent \footnotesize{johannes.knoerzer@mpq.mpg.de}}
\thanks{\newline \noindent \footnotesize{$\dagger$ Present address: Amazon Web Services, New York. This work was done prior to joining Amazon Web Services.}}
\thanks{\newline \noindent \footnotesize{$\ddagger$ Present address: IMEC, Remisebosweg 1, 3001 Leuven, Belgium.}}

\date{\today}

\begin{abstract}

Wigner crystals are prime candidates for the realization of regular electron lattices under minimal requirements on external control and electronics.
However, several technical challenges have prevented their detailed experimental investigation and applications to date.
We propose an implementation of two-dimensional electron lattices for quantum simulation of Ising spin systems based on self-assembled Wigner crystals in transition-metal dichalcogenides.
We show that these semiconductors allow for minimally invasive all-optical detection schemes of charge ordering and total spin.
For incident light with optimally chosen beam parameters and polarization, we predict a strong dependence of the transmitted and reflected signals on the underlying lattice periodicity, thus revealing the charge order inherent in Wigner crystals.
At the same time, the selection rules in transition-metal dichalcogenides provide direct access to the spin degree of freedom via Faraday rotation measurements.
 
\end{abstract}
\maketitle

\section{Introduction \label{sec:introduction}}

Ever since its theoretical inception 85 years ago \cite{wigner34},
Wigner crystallization has stimulated both theoretical and experimental research to find unambiguous evidence for this elusive state of matter.
Since the earliest indication for quantum Wigner crystals (WCs) obtained from high-magnetic-field transport measurements \cite{mendez83,andrei88}, it has proven to be a very demanding task to study WCs, especially in a minimally invasive manner without destroying the crystalline order.
Recent experimental work demonstrated non-destructive read-out of the charge distribution of one-dimensional WCs in carbon nanotubes \cite{shapir18}.
However, it remains an open challenge to find approaches for the non-invasive detection of WCs in two-dimensional and a broader range of one-dimensional quantum systems.

Apart from a fundamental interest in the physics of Wigner crystallization, self-assembled crystals promise a route towards highly ordered and scalable many-body systems under minimal external control.
Thus, they meet some of the key requirements posed by quantum computers \cite{divincenzo00} and simulators \cite{cirac12}.
It has therefore been proposed that Wigner crystals hosted in semiconductor
nanostructures \cite{deshpande08,antonio15}, trapped above the surface of liquid helium \cite{platzman99,dykman00}
or composed of trapped ions \cite{porras06,biercuk09}
can be utilized for quantum information processing and simulation.
In particular, electrons confined to low-dimensional semiconductors \cite{silvestrov17} may be brought into the low-temperature regime $k_\mathrm{B} T \ll \varepsilon_\mathrm{F}$ (Fermi energy $\varepsilon_\mathrm{F}$) where quantum phenomena occur and spin-exchange interactions can play an important role.
Since solid-state systems also offer a genuine prospect for miniaturization and on-chip integration, the quest for a faithful implementation of solid-state quantum WCs at zero magnetic field remains tantalizing.

\begin{figure}[b!]
\centering
\begin{overpic}[width=0.8\linewidth]{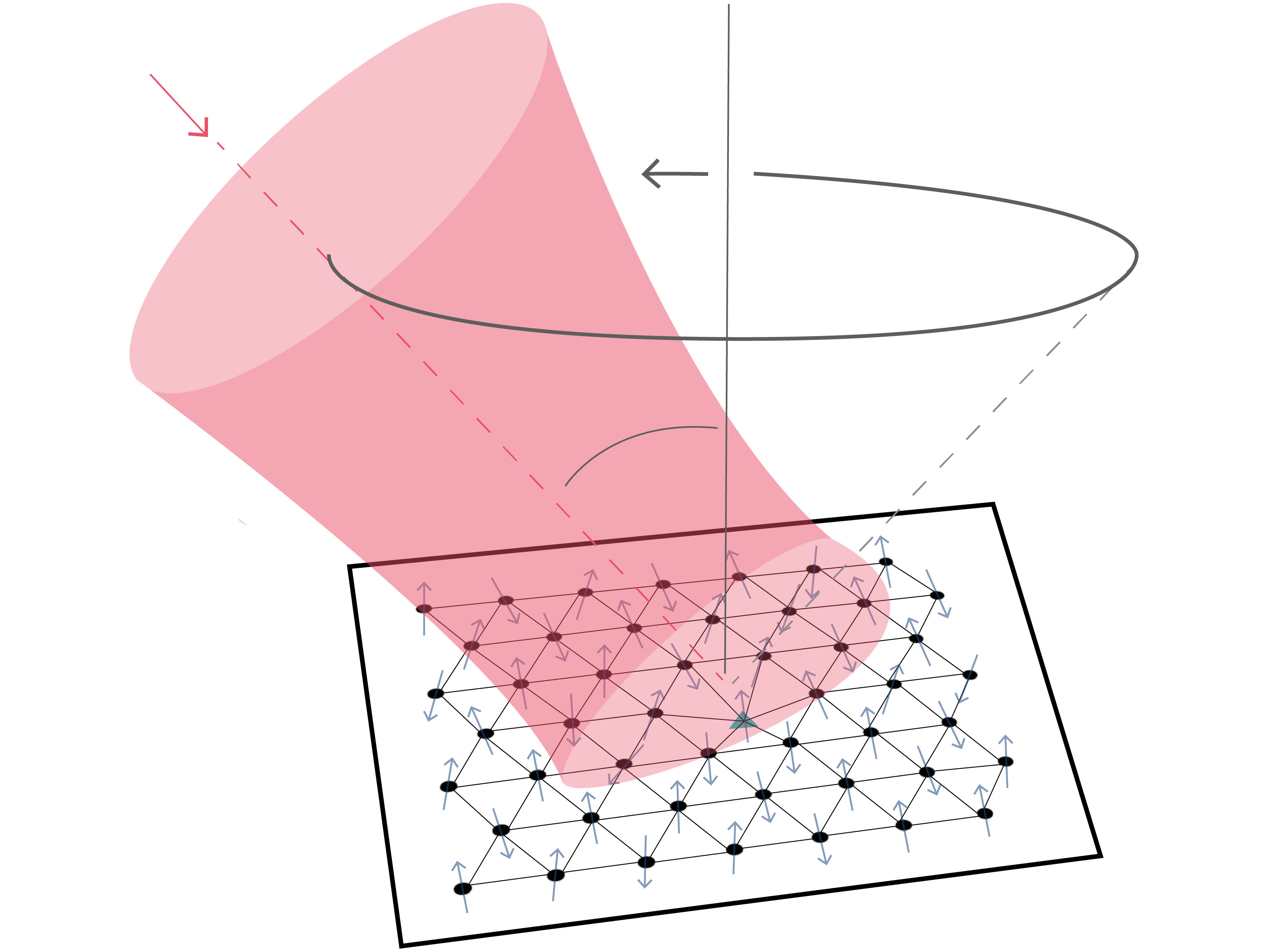}
\put(0,0){
\begin{tikzpicture}
\path (0,0) node (x) {}
         (0.5,0.5) node[circle,draw,color=white](y){1};
\draw[white] (x) -- (y);
\node (k) at (0.52,4.2) {\textcolor{red}{$\vec{k}$}};
\node (theta) at (3.4,2.3)
{\textcolor{black}{$\theta$}};
\node (phi) at (6.1,4.0)
{\textcolor{black}{$\phi$}};
\node[rotate=7] (TMD) at (5.3,0.0) {\textcolor{gray}{TMD}};
\end{tikzpicture}}
\end{overpic}
\caption{
(color online). Schematic illustration of proposed setup and optical detection scheme.
Charge ordering of electrons in a lattice (black dots) competes with random disorder-induced dislocations of lattice sites in the presence of impurities and defects (green triangle).
The angle-dependent ($\phi$) reflection of a tilted ($\theta$) focused laser beam with wavevector $\vec{k}$ from a WC probes its lattice geometry.
Light polarization provides further information about the spin via optical selection rules of TMDs. 
}
\label{fig:schematic-figure}
\end{figure}
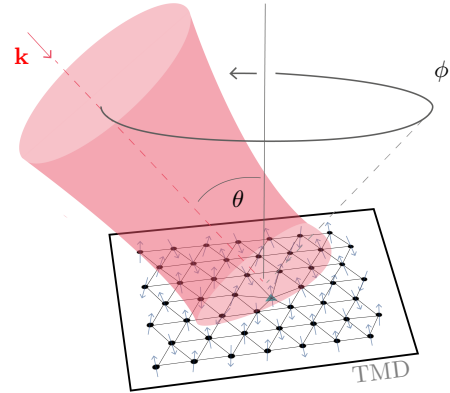

As recently pointed out, monolayer transition-metal dichalcogenides (TMDs) \cite{zarenia17} and TMD-based moir\'e superlattices \cite{regan19,tang19,shimazaki19} are unique platforms for realizing strongly correlated systems and the study of WCs in particular owing to the combination of reduced screening in two dimensions and a relatively high effective electron mass.
Their optical bandgap offers exciting possibilities to probe quasiparticle excitations, e.g., excitons or trions \cite{berkelbach13,mak13,chernikov14,wang18} optically \cite{zeytinoglu17,back18,scuri18}.

In this work, we demonstrate the potential of scalable quantum simulators based on two-dimensional WCs in TMDs and propose an all-optical detection scheme for charge ordering and partial spin information in these systems (see Fig.~\ref{fig:schematic-figure}).
In particular, the scheme possesses three key properties:
(i) It provides clear evidence for Wigner crystallization in monolayer TMDs.
(ii) Under conditions specified below, the detection scheme is non-invasive and leaves charge and spin order intact.
(iii) Optical selection rules provide spin-selective addressability which is a crucial requirement for quantum simulation.

\section{Theoretical Framework \label{sec:theory}}

\textit{Wigner crystals.}\textemdash
At electron densities $n$ below a critical density $n_\mathrm{cr}$
and in the presence of an external confinement
potential, interacting charge carriers (refered to as electrons in the following)
arrange themselves in a lattice
\footnote{For simplicity, we refer to this lattice as a WC, despite the absence of long-range order. In the literature, these systems are also referred to as Wigner molecules.},
leading to a periodic modulation of charge density $n(\mathbf{r})$.
In this low-density regime, electrostatic interactions dominate over the kinetic
energy of electrons.
In two dimensions, this regime is characterized by a sufficiently large
interaction parameter $r_\mathrm{s} = 1/(\sqrt{\pi n}a_\mathrm{B})$,
with the Bohr radius $a_\mathrm{B} = 4\pi\varepsilon \hbar^2 / (e^2 m)$,
effective electron mass $m$ and permittivity $\varepsilon$.
Monolayer TMDs feature an extraordinarily small Bohr radius $a_\mathrm{B} \gtrsim 0.5~$nm and thus render the large-$r_\mathrm{s}$ regime accessible at experimentally achievable \cite{mak14,roch19} densities $n \lesssim n_\mathrm{cr}$.
For our calculations, we choose $n_\mathrm{cr} = 10^{11} \ \mathrm{cm}^{-2}$ \cite{zarenia17} and $m = 0.5m_\mathrm{0}$ (representative of $\mathrm{MoX}_2$ monolayers where $\mathrm{X} = \mathrm{S}, ~ \mathrm{Se}$ \cite{kumar12}), where $m_0$ denotes the bare electron mass.
In a square (triangular) lattice, this maximum electron density corresponds to a minimum lattice spacing of $a \gtrsim 32~\mathrm{nm}$ ( $a \gtrsim 34~\mathrm{nm}$).

\textit{Model.}\textemdash
We consider $N$ electrons trapped at $z = 0$ in a global harmonic potential such that the total potential reads
\begin{equation}\label{eq:hamilt1}
V(\mathbf{r}_1, ..., \mathbf{r}_N) = \frac{m \omega^2}{2}\sum_{i=1}^N \left ( x_i^2 + y_i^2 \right ) + \sum_{i\neq j} V_\mathrm{int} \left ( \mathbf{r}_i, \mathbf{r}_j \right ),
\end{equation}
where $\mathbf{r}_i = ( x_i, \ y_i, \ 0)$ denotes the position of the $i$th electron.
The confinement is characterized by the trapping frequency $\omega$ and $V_\mathrm{int}$ denotes the two-body interaction potential.
In TMDs, the former may be induced by strain \cite{roldan15,amorim16} or defined via local gates \cite{wang18NatNano} and
the latter is usually modeled by the Keldysh potential \cite{keldysh79},
\begin{equation}\label{eq:keldysh-interaction}
V_\mathrm{int}(\mathbf{r}_i,\mathbf{r}_j) = \frac{\pi e^2}{2r_0} \left [ H_0 \left (\frac{|\mathbf{r}_i-\mathbf{r}_j|}{r_0} \right ) - Y_0 \left (\frac{|\mathbf{r}_i-\mathbf{r}_j|}{r_0} \right) \right ],
\end{equation}
with a material-specific length scale $r_0 \approx 5~\mathrm{nm}$.
$H_0$ and $Y_0$ are Struve and Bessel functions, respectively.
At electron concentrations $n < n_\mathrm{cr}$, the inter-particle distance $|\mathbf{r}_i-\mathbf{r}_j| \gg r_0$ and hence $V_\mathrm{int}(\mathbf{r}_i,\mathbf{r}_j) \sim 1/|\mathbf{r}_i-\mathbf{r}_j|$ behaves like a Coulomb potential.

\begin{figure}[b!]
   \begin{minipage}[b]{1.0\linewidth}
   \begin{overpic}[width=1.0\linewidth]{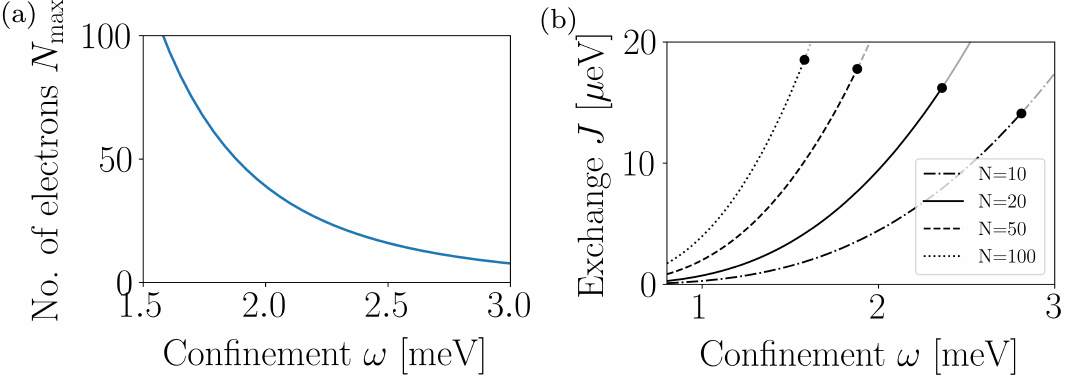}
   \end{overpic}
   \end{minipage}
\caption{
Spin coupling and system size.
(a)
Maximum number of electrons as a function of $\omega$ such that $n < n_\mathrm{cr}$.
(b)
Coupling constant $J$ as a function of
the confinement $\omega$ for different particle numbers $N = 10$ (\textit{dash-dotted}), $N = 20$ (\textit{solid}), $N = 50$ (\textit{dashed}), $N = 100$ (\textit{dotted}).
Black dots: maximum frequency $\omega$ for given $N$ such that $n < n_\mathrm{cr}$.
}
\label{fig:exchange-constant}
\end{figure}

In a WC, the electrons are localized around lattice sites at $\mathbf{r}^0_i$ ($i = 1, ..., N$) which can be determined from the equilibrium conditions $\nabla_i V \vert_{\mathbf{r}_i = \mathbf{r}_i^0} = 0$.
Numerical calculations show that harmonic confinement potentials, as described in Eq.~\eqref{eq:hamilt1}, give rise to triangular lattice geometries while other potentials can give rise to, e.g., square lattices; see Appendix \ref{app:modes} for details.
For any $\omega$, the maximum number of WC electrons can be calculated given a critical density, and vice versa.
Small systems containing $N \sim (10 - 100)$ electrons require $\hbar \omega \sim (1 - 3)~$meV at $n \sim n_\mathrm{cr}$ (see Fig.~\ref{fig:exchange-constant}).

The strong interactions in Eq.~\eqref{eq:keldysh-interaction} enable the description of charge excitations in terms of phonons in the WC.
These can be expressed as small displacements $\mathbf{q}_i = \mathbf{r}_i - \mathbf{r}^0_i$ ($i = 1, ..., N$) from the lattice sites such that $V = (m/2) \sum \mathcal{K}_{ij}^{\alpha\beta} q_i^\alpha q_j^\beta$ ($\alpha, \beta \in \{ x, y \}$) with an elasticity matrix $\mathcal{K}$.
All $2N$ normal modes of the system with eigenfrequencies $\Omega_n$ ($n = 1, ..., 2N$) are readily obtained by diagonalization of $\mathcal{K}$ and for the non-zero eigenfrequencies one finds that $\Omega_n \gtrsim \omega$ (cf.~Appendix \ref{app:modes}).
Given the relation between $\omega$ and $N$ at $n \sim n_\mathrm{cr}$, this indicates that large WCs have low-energy phonon modes.
Using anharmonic potentials, there is no limit placed on $N$ by the phonon modes or $n_\mathrm{cr}$.

\textit{Requirements.}\textemdash
Wigner crystallization requires low disorder.
Disorder-induced potential fluctuations are incorporated based on Eq.~\eqref{eq:hamilt1} by adding further randomly distributed local confinement terms
to analyze the impact of impurities (e.g., atomic defects or charges) on the electron lattice.
In order to obtain a regular lattice structure with an approximately equidistant spacing between adjacent electrons (see schematic Fig.~\ref{fig:schematic-figure}), the impurity density $n_\mathrm{imp}$ should be significantly smaller than the electron density, i.e.~$n_\mathrm{imp} \lesssim 0.1 n$; see Appendix \ref{app:disorder} for details.
To date, atomic and charge defects in TMDs still prevent the realization of systems with sufficiently low disorder \cite{rhodes19}.
However, both sample quality and deterministic control over defects \cite{klein19} have been improving rapidly in recent years and defect densities around $n_\mathrm{cr}$ can already be achieved.
Moreover, WCs require sufficiently low temperature.
Cooling into the motional ground state requires low temperatures $T \sim 1~\mathrm{K}$ for $\hbar \omega \lesssim \mathrm{meV}$,
as the thermal occupation $\bar n_\mathrm{th} = 1/[\exp(\hbar\Omega_n/(k_\mathrm{B}T))-1]$ of the modes increases as $\omega$ is decreased (cf.~Appendix \ref{app:temperature}).

There are many interesting aspects about the dynamics of strongly correlated electrons that can be studied in the system we describe, including the entanglement properties of the ground state, the nature and dynamics of excitations and the transitions to neighboring phases.
In the following, we focus on the spin physics.

\section{Spin physics \label{sec:spins}}

TMD monolayers exhibit strong spin-orbit coupling and an intricate interplay between spin and valley degrees of freedom.
Here we focus on the case where, by energetic isolation of the lower spin states of the conduction band, spin and valley become locked \cite{dey17}.
For this reason, we require that the electron density be sufficiently low such that the Coulomb interaction energy $E_\mathrm{int} \sim r_\mathrm{s} \cdot \varepsilon_\mathrm{F} = r_\mathrm{s} \pi \hbar^2 n/m$ is small compared to the spin-orbit splitting in the conduction band, $\Delta_\mathrm{SO}^c$. At $n \lesssim n_\mathrm{cr}$, one typically finds $E_\mathrm{int} \lesssim 10~$meV, such that the above condition is readily satisfied in MoSe$_2$ ($\Delta_\mathrm{SO}^c \approx 23~$meV), though not necessarily in MoS$_2$ ($\Delta_\mathrm{SO}^c \approx 3~$meV) \cite{kormanyos14}.
Nevertheless, the requirement can be met in all TMDs by considering holes instead of electrons, since the spin-orbit splitting in the valence band $\Delta_\mathrm{SO}^v$ is on the order of a hundred meV \cite{wang18}.

At low temperature and small displacements $\mathbf{q}_i$, we assume that the electron spins are localized around the lattice sites at $\mathbf{r}_i^0$.
Adjacent spins are coupled via exchange interactions that can be either ferromagnetic or antiferromagnetic, depending on the density $n$ \cite{thouless65,drummond09}.
Here, we provide an estimate for the magnitude of the spin-spin coupling, demonstrating the potential of TMD-based electron lattices as a platform for quantum simulation of prototypical spin systems.
As exchange couplings decay exponentially with $a^2$, where $a$ denotes the inter-particle distance, the low-density regime necessary for WCs stands in contrast with the strong couplings of interest for spin physics.
However, at intermediate densities $n \lesssim n_\mathrm{cr}$ we still find significant exchange couplings which exceed predicted spin relaxation rates \cite{wu16,pearce17}.

Due to the spin polarization in each of the $K$ and $K^\prime$ valleys, we find that the effective spin model in the spin-valley locked, low-temperature regime reduces to an Ising Hamiltonian (cf.~Appendix \ref{app:exchange-coupling} for details) of the form
\begin{equation}\label{eq:J-main-text}
H_\sigma = \sum_{i, j} J_{ij} \sigma_i^z \sigma_j^z.
\end{equation}
Here $\sigma_i^z$ is a Pauli operator and $J_{ij}$ denotes the coupling strength between spins at sites $i$ and $j$.
In a tight-binding approximation, we calculate $J_{ij}$ ($1 \leq i < j \leq N$) using Gaussian ansatz wavefunctions centered around the sites $\mathbf{r}_i^0$.
The width of these wave functions is expressed in terms of the normal mode frequencies $\Omega_n$ and, upon inserting typical material parameters, we find for the magnitude $J$ of the spin-spin interaction between nearest neighbours typical values in the range $J \sim (5 - 30)~\mu$eV for $n \lesssim n_\mathrm{cr}$.
Due to the exponential decay of $J_{ij}$ with distance, nearest-neighbour interactions are dominant and typically roughly one order of magnitude larger than next-nearest-neighbour interactions.
In Fig.~\ref{fig:exchange-constant}(b), we show the resulting spin-coupling constant $J$ as a function of $\omega$ for different particle numbers $10 \leq N \leq 100$.
At the intermediate densities $n \lesssim n_\mathrm{cr}$ considered here, we find antiferromagnetic exchange couplings which can result in geometrical frustration \cite{moessner01} depending on the lattice structure.

\begin{figure}[t!]
\centering
\includegraphics[width=1\linewidth]{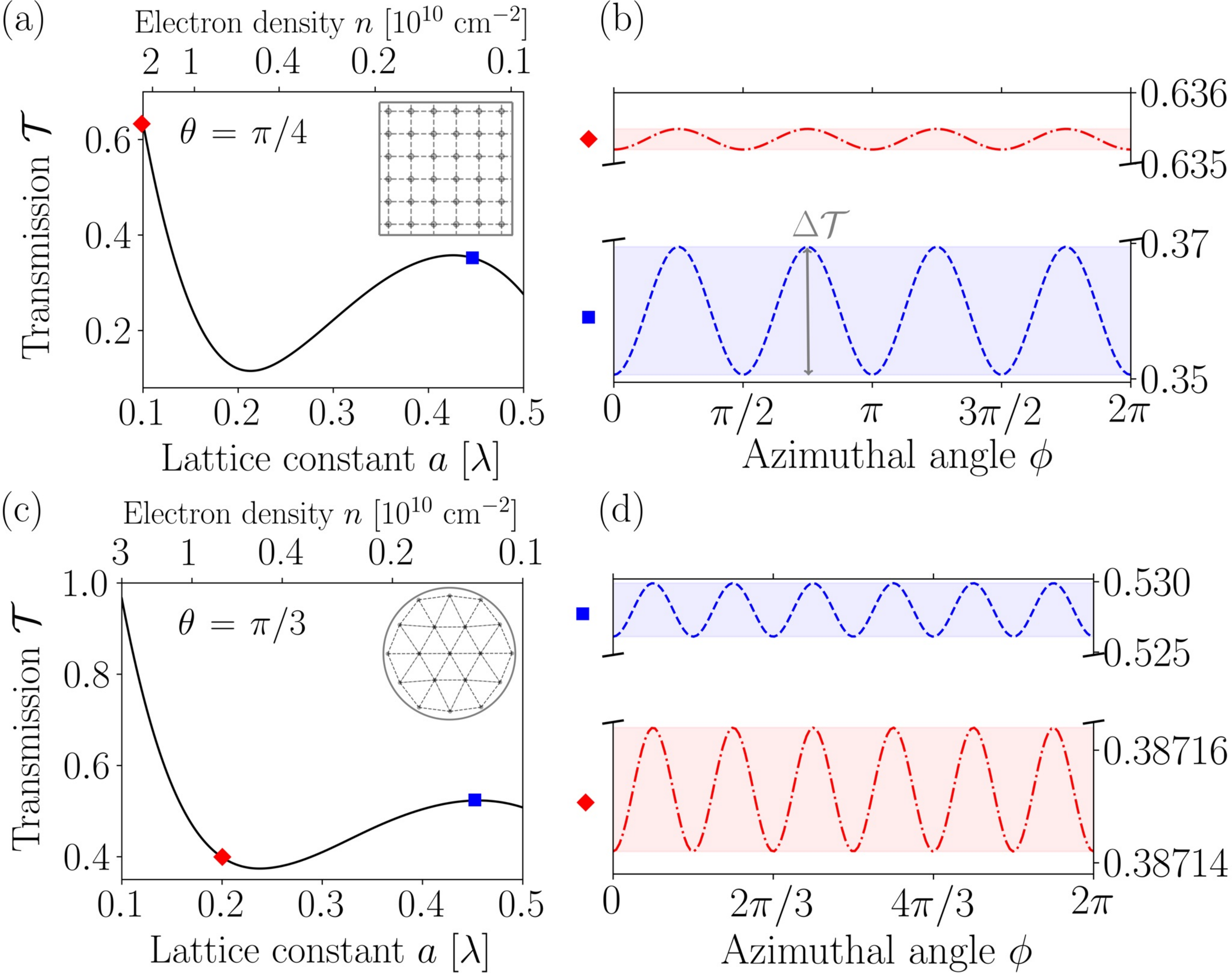}
\caption{
Density and angle-dependent transmission of elliptically polarized (see Eq.~\eqref{eq:ellip-pol}) incoming light beam at incident angle $\theta$ and in-plane rotation angle $\phi$ (see Fig.~\ref{fig:schematic-figure}).
(a) Transmission $\mathcal{T}$ at a tilt angle $\theta = \pi/4$ for a square lattice configuration as a function of lattice constant $a$ and density $n$.
(b) $\mathcal{T}(\phi)$ at chosen values for $a/\lambda = 0.1$ (\textit{red, dash-dotted line}), $a/\lambda = 0.45$ (\textit{blue, dashed line}) and same parameters as in (a).
Contrast $\Delta \mathcal{T}$ is depicted by oscillation amplitude of $\mathcal{T}(\phi)$.
(c) Same as (a) but for a triangular lattice configuration. Angle of incidence $\theta = \pi/3$.
(d) $\mathcal{T}(\phi)$ at chosen values for $a/\lambda = 0.2$ (\textit{red, dash-dotted line}), $a/\lambda = 0.45$ (\textit{blue, dashed line}) and same parameters as in (c).
\textit{Numerical parameters:}
Gaussian beam waist $w_0 = 1.0 \lambda$, $N = 40 \times 40$, detuning $\Delta_0 = 0$.
}
\label{fig:transmission-readout}
\end{figure}

\section{Optical readout \label{sec:readout}}

We now address the optical detection of charge ordering in TMD-based WCs and
consider an incoming ($z < 0$) Gaussian laser beam $\mathbf{E}_\mathrm{in}(\mathbf{r})$ with wavelength $\lambda$ focused to a spot on the electron lattice ($z = 0$) at a tilt angle $\theta$ (see Fig.~\ref{fig:schematic-figure}).
Our approach is similar in nature to the one taken in Refs.~\cite{bettles16,shahmoon17}, where the reflection and transmission of arrays of discrete atomic emitters in a lattice configuration was analyzed.
Such an approach is valid for highly localized charges \cite{thomas61}, in contrast to the study of mobile polarons \cite{sidler17}.
Due to optical transition selection rules in monolayer TMDs, specific electron spin states can be addressed using circularly polarized $\sigma^+$ and $\sigma^-$ light.
For example, $\sigma^-$ ($\sigma^+$) light may couple a WC electron in a $\ket{\uparrow_\mathrm{\small K}}$ ($\ket{\downarrow_{\mathrm{\small K}^\prime}}$) spin state to a trionic state $\ket{ \uparrow_{\mathrm{\small K}},  \downarrow_{\mathrm{\small K}^\prime} \Downarrow_{\mathrm{\small K}^\prime} }$ ($\ket{ \downarrow_{\mathrm{\small K}^\prime}, \uparrow_{\mathrm{\small K}} \Uparrow_{\mathrm{\small K}} }$) with a hole spin $\Uparrow$ ($\Downarrow$) in the $K^\prime$ ($K$) valley.
For our calculations, we assume a low-amplitude light beam with sufficiently small detuning $\hbar \Delta_0 \ll E_b, E_g$ from the trion resonance such that other quasiparticle excitations and transitions can be neglected.
Prototypical values for the trion binding energy $E_b \sim 20~$meV and quasiparticle band gap $E_g \sim 500~$meV are given in Ref.~\cite{wang18}.
When the incoming beam is sufficiently close to resonance with a dipole transition at lattice points $\mathbf{r}_n^0$, the scattered light field $\mathbf{E}(\mathbf{r})$ at position $\mathbf{r}$ is obtained by solving a set of coupled linear equations,
\begin{equation}\label{eq:lip-schwinger}
\mathbf{E}(\mathbf{r}) = \mathbf{E}_{\mathrm{in}}(\mathbf{r}) + \frac{4\pi^2}{\varepsilon_0 \lambda^2} \sum_{n=1}^N G(k,\mathbf{r}, \mathbf{r}_n^0) \alpha_n(\Delta_0) \mathbf{E}(\mathbf{r}_n^0),
\end{equation}
with the detuning from resonance $\Delta_0$, the dyadic Green's function $G$ evaluated at $k=2\pi/\lambda$ and the polarizability tensor $\alpha_n$.
The magnitude of the polarizability tensor is given by the scalar polarizability $\alpha(\Delta_0)$, while the orientation depends on the electron spin at site $n$; see Appendix \ref{app:optical-readout} for more details.

In order to probe charge ordering, it is advantageous to address all WC electrons equally.
To this end, we assume for the following discussion that the WC is fully spin polarized, which could be achieved by applying a large magnetic field or via optical pumping \cite{yang15}.
Alternatively, one could consider a TMD heterobilayer system where an electron-hole pair excited in one layer forms a trion state with a WC electron in the other layer, such that both valleys can be addressed independent of the spin of the resident electron \cite{wang18}.

The total power $P$ transmitted by the WC to $z > 0$ is obtained by integrating the transmitted signal ($\mu_0 = 1$),
\begin{equation}\label{eq:power}
P = \frac{1}{2} \int_\mathcal{S} \mathrm{Re}\left [ \mathbf{E} \times \mathbf{B}^* \right ] \cdot \hat{\mathbf{z}} \ \mathrm{d} \mathrm{A},
\end{equation}
with the electric and magnetic fields $\mathbf{E}$ and $\mathbf{B}$, respectively, and $\mathbf{B}^*$ denotes the complex conjugate of $\mathbf{B}$.
The transmission $\mathcal{T} = P_\mathrm{wc}/P_0$ is calculated as a function of density $n$, incidence angle $\theta$, and rotation angle $\phi$ (see Fig.~\ref{fig:schematic-figure}) by comparing the transmitted power $P_\mathrm{wc}$ in the presence of a WC with a reference signal $P_0$ obtained in the absence of localized dipoles \cite{bettles16}, e.g.~in a system with no doping at $n=0$.

In Fig.~\ref{fig:transmission-readout}, $\mathcal{T}$ is shown as a function of the electron density $n \sim 1/a^2$ for square [Fig.~\ref{fig:transmission-readout}(a)] and triangular [Fig.~\ref{fig:transmission-readout}(c)] lattices with a lattice constant $a$.
Here we consider $\Delta_0 = 0$, which corresponds to a wavelength $\lambda \sim (700 - 800)~$nm  in state-of-the-art TMD setups \cite{singh16,courtade17}.
We choose $\theta$ such that the cross section of the Gaussian beam is small enough and does not exceed the size of the WC.
Varying the twist angle $\phi$ of the laser beam leads to smooth variations in $\mathcal{T}(\phi)$.
The periodic modulation of $\mathcal{T}(\phi)$ reflects the rotational symmetry of the WC.
Figs.~\ref{fig:transmission-readout}(b) and (d) display the $2\pi/4$ and $2\pi/6$ rotational symmetry of a square and triangular lattice, respectively.
The amplitude of this periodic signal shows that the contrast $\Delta \mathcal{T} = \underset{0 \leq \phi< 2\pi}{\mathrm{max}} \mathcal{T}(\phi) -\underset{0 \leq \phi< 2\pi}{\mathrm{min}} \mathcal{T}(\phi)$ can be of the order of a few percent.
This modulation provides an unambiguous experimental signature of Wigner crystallization.
The beam parameters and polarization of the incident light can be optimized to maximize the transmission contrast (cf.~Appendix \ref{app:optical-readout}).
Momentum transfer onto the WC can be safely neglected since the recoil energy $E_\mathrm{R} = \hbar^2 k^2 / (2m) \sim (5 - 10) \mu\mathrm{eV}$ is much smaller than interaction energy and trapping potential.
This approach already incorporates spin information, as it can be used to detect ferromagnetic ground states and may pick up signatures of the lattice constant $2a$ prevailing in an antiferromagnetic ground state.

\begin{figure}[t!]
\centering
\includegraphics[width=1\linewidth]{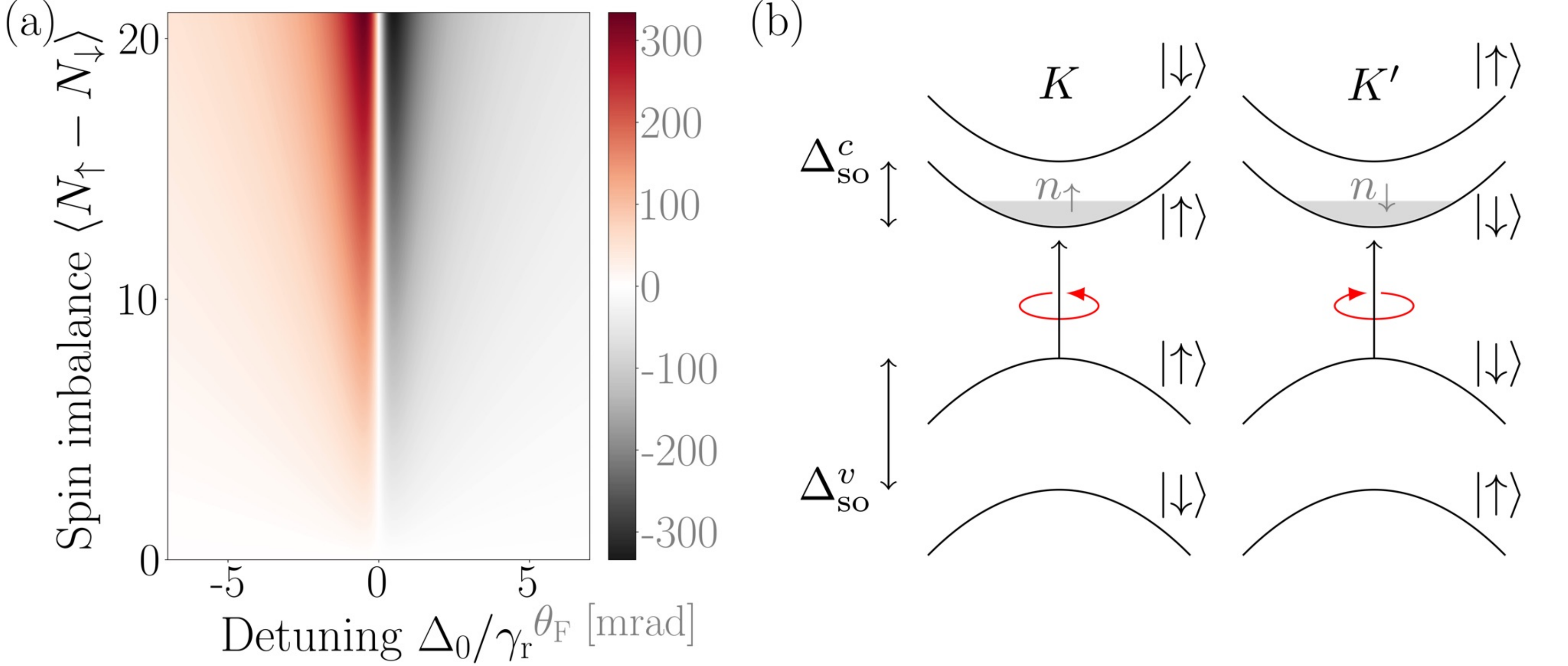}
\caption{
Faraday rotation and optical selection rules.
(a) $\theta_\mathrm{F}$ from Eq.~\eqref{eq:faraday} as a function of detuning $\Delta_0$ from the bare resonance and spin imbalance $N_\uparrow - N_\downarrow$.
Results for a total number of $N = N_\uparrow + N_\downarrow = 25$ electrons in a square lattice at $a/\lambda = 0.4$.
(b) Energy level diagrams for conduction and valence bands at the $K$ and $K^\prime$ valleys of $\mathrm{MoX}_2$ monolayers with spin-orbit splittings between $\ket{\downarrow}$ and $\ket{\uparrow}$ in the conduction ($\Delta_\mathrm{SO}^c$) and valence ($\Delta_\mathrm{SO}^v$) bands.
Carrier densities $n_\uparrow$ and $n_\downarrow$ in the $\ket{\uparrow_{\mathrm{\small K}}}$ and $\ket{\downarrow_{\mathrm{\small K}^\prime}}$ conduction bands, respectively.
Right-circularly (left-circularly) polarized light couples only to spin-up (spin-down) electron states in the $K$ ($K^\prime$) valley.
\textit{Numerical parameters:} nonradiative linewidth $\hbar \gamma_\mathrm{nr} = 0$, tilt angle $\theta = 0$ (normal incidence) and beam waist $w_0 = 1.0\lambda$.
}
\label{fig:faraday-readout}
\end{figure}

\textit{Faraday rotation}.\textemdash
While we have focused on the detection of charge ordering in a spin-polarized WC before, we now further examine the spin degree of freedom by analyzing the polarization of the scattered field.
With the probe beam $\mathbf{E}_\mathrm{in}$ detuned far enough from the trionic resonance, the presence of the optical transition merely imprints a state-dependent phase shift on the incoming field.
According to selection rules of monolayer TMDs \cite{xiao12,xu14}, $\sigma^+$ ($\sigma^-$) polarized light couples to the resident electron density $n_\uparrow$ ($n_\downarrow$) in the $K$ ($K^\prime$) valley (see Fig.~\ref{fig:faraday-readout}(b)).
In optical Faraday (Kerr) rotation using linearly polarized light, the polarization of the transmitted (reflected) part of the light is rotated by an angle $\theta_\mathrm{F}$ which depends on the spin imbalance $n_\uparrow$ - $n_\downarrow$ \cite{berezovsky06,yang15}.
Here we inspect the Faraday rotation of an incident $s$ or $p$-polarized beam, which is given by \cite{wolf}
\begin{equation}\label{eq:faraday}
\theta_\mathrm{F} = \frac{1}{2} \arctan \frac{2\mathrm{Re}\chi_\mathrm{F}}{1 - |\chi_\mathrm{F}|^2},
\end{equation}
where $\chi_\mathrm{F} = t_{ps}/t_{ss}$ for $s$-polarized light ($\chi_\mathrm{F} = -t_{sp}/t_{pp}$ for $p$-polarized light) depends on the Jones matrix elements $t_{ss}$, $t_{ps}$ ($t_{pp}$, $t_{sp}$) encoding the polarization state of the scattered light \cite{visnovsky}.
We consider $N_\uparrow$ ($N_\downarrow$) electrons in the $\ket{\uparrow_\mathrm{K}}$ ($\ket{\downarrow_{\mathrm{K}^\prime}}$) conduction band and numerically calculate $\theta_\mathrm{F}$ as a function of spin imbalance $N_\uparrow - N_\downarrow$ and detuning $\Delta_0$.
Here we assume that the electron sites $\mathbf{r}_i^0$ are distributed in a square-lattice configuration in the spot of the beam with $N_\uparrow$ ($N_\downarrow$) randomly assigned $\ket{\uparrow}$ ($\ket{\downarrow}$) states.
We average over many such configurations.
In Fig.~\ref{fig:faraday-readout}(a) the resulting Faraday rotation is depicted for a $p$-polarized input field, yielding the strongest signal at $|\Delta_0| = \gamma_\mathrm{r}/2$ with the radiative linewidth $\gamma_\mathrm{r}$.
For the strongly localized quantum emitters considered here, we estimate $\hbar \gamma_\mathrm{r} \sim 10^{-2}~\mu$eV.
Nonradiative decay processes can also be taken into account in our framework, yielding weaker Faraday signals for larger nonradiative decay rates $\gamma_\mathrm{nr}$ (cf.~Appendix \ref{app:optical-readout}).
Since the Faraday rotation is proportional to $N_\uparrow - N_\downarrow$, it provides a measure for the spin imbalance in the system.
With this tool, one may distinguish between ferromagnetic and antiferromagnetic configurations or even locally probe domain walls in the spin system, where the spatial resolution would be limited by the spot size $\sim \lambda^2$.

\section{Summary \& Outlook \label{sec:outlook}}

In conclusion, we have proposed an all-optical detection scheme for TMD-based Wigner crystals, highlighting their potential as a platform for the quantum simulation of geometrically frustrated magnetism with adjustable and self-assembled lattice structures.
Beyond the Ising model considered here, richer spin physics with multi-spin exchange interactions has been predicted for these systems, potentially offering a platform to study three- and four-body interactions \cite{voelker01,matveev14}.
Moreover, recent results show that multi-electron quantum dots hold promise as exchange-based mediators of quantum information \cite{malinowski19}.
In this context, intermediate-scale Wigner crystals in 2D semiconductors could be interesting for achieving long-range spin coupling with minimal external control requirements \cite{antonio15}.
Control over the spin degree of freedom may be provided via magnetic fields or optical pumping into a specific valley, e.g.~in parts of the system to study the formation of domain walls.
Inversion symmetric TMD bilayers, whose bands are spin degenerate, may further give rise to a wider range of spin Hamiltonians and allow for coherent optical control of the electron spin as no momentum is required to flip the spin.
High-quality samples of monolayer TMDs should provide access to first proof-of-principle experiments with small system sizes.
Local spin probes may be enabled by illuminating only parts of the WC.
Besides the optical techniques we propose, which we believe can be readily implemented given sufficiently clean samples, we envisage that it might become possible in the future to extend existing and developing work on high-resolution electron beam imaging with (close-to) single site resolution
\cite{jiang18,yoo19} to the point that a single electron charge can be directly spatially probed.
Furthermore, other detection schemes could be considered like magnetic noise spectroscopy \cite{agarwal17}, microwave spectroscopy \cite{oosterkamp98}, or using surface acoustic waves in piezoelectric TMD monolayers \cite{rezk16}.

\begin{acknowledgments}
We acknowledge valuable discussions with Matteo Barbone, Deung-Jang Choi, Jonathan Finley, Alexander Holleitner, Malte Kremser, \"Ors Legeza, and Jonah Waissman.
JK, RS and JIC acknowledge support from the DFG (German Research Foundation) under Germany's Excellence Strategy - EXC-2111 - 39081486.
GG acknowledges support by the Spanish \emph{Ministerio de Ciencia, Innovation y Universidades} through the Project No. 2017-83780-P, and the European FET-OPEN project SPRING ($\#863098$).
We also acknowledge the Max Planck Harvard Research Center for Quantum Optics for support.
JK and JIC thank Harvard University for hospitality during several visits.
\end{acknowledgments}

\appendix
\section{Calculation of lattice structure and normal modes \label{app:modes}}
We consider a general potential of the form
\begin{equation}\label{eq:pot-energy-2d-appendix}
V_p = \sum_{i=1}^N \mu_p \left ( x_i^p + y_i^p \right ) + \sum_{i\neq j} V_\mathrm{int}(\mathbf{r}_i,\mathbf{r}_j),
\end{equation}
where $\mu_p$ is the strength of the potential and the interaction potential $V_\mathrm{int}$ is modeled by the Keldysh interaction
potential given in Eq.~(2).
The results presented in the main text are derived for the special case $p = 2$ and $\mu_2 = m \omega^2 / 2$.

\begin{figure}[t!]
   \begin{minipage}[b]{1.0\linewidth}
   \begin{overpic}[width=1.0\linewidth]{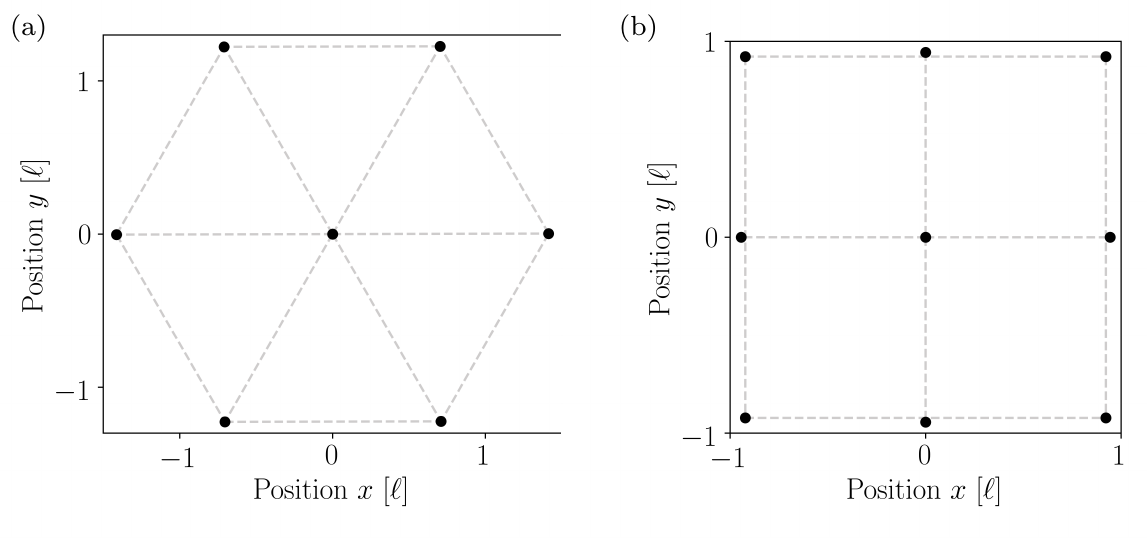}
   \end{overpic}
   \end{minipage}
\caption{
Lattice configurations $\{ \mathbf{r}_i^0 \}_{1\leq i\leq N}$ (\textit{black dots}) for small
systems of (a) $N = 7$ electrons in a harmonic potential with $p = 2$
and (b) $N = 9$ electrons in an anharmonic potential with $p = 8$.
}
\label{fig:geometries}
\end{figure}

\subsection{Lattice structure}
The lattice sites $\mathbf{r}_i^0$ are calculated by solving the equations
\begin{equation}
    \frac{\partial V_p}{\partial x_i}\bigg\vert_{\mathbf{r}_i=\mathbf{r}_i^0} = \frac{\partial V_p}{\partial y_i}\bigg\vert_{\mathbf{r}_i=\mathbf{r}_i^0} = 0
\end{equation}
for each electron $i \in \{ 1, ..., N \}$.
This leads to a set of $2N$ coupled equations which are of the form
\begin{equation}\label{eq:coupled-eqs-ri0}
    \mu_p p\alpha_i^{p-1} + \xi \sum_{j \neq i} \left ( \alpha_j - \alpha_i \right ) h\left ( |\mathbf{r}_j - \mathbf{r}_i|/r_0 \right ) = 0,
\end{equation}
with $\alpha \in \{ x, y \}$, $\xi = \pi e^2/(2 r_0^3)$ and the function
\begin{equation}\label{eq:functions-g-and-h}
h(x) = H_{-1}(x)-H_{1}(x) + Y_1(x) - Y_{-1}(x) + \frac{1}{\sqrt{\pi} \Gamma(\frac{3}{2})},
\end{equation}
which is obtained by making use of recurrence relations for the \textit{Struve} and \textit{Bessel functions of the second kind} $H_\nu$ and $Y_\nu$ ($\nu \in \mathbb{N}$), respectively.
In order to solve Eqs.~\eqref{eq:coupled-eqs-ri0}, it is instructive to introduce dimensionless variables scaled by a length scale $\ell = [e^2/(4\pi \varepsilon p \mu_p)]^{1/(p+1)}$.
For $r_0\ll\ell$, we find that the obtained lattice configurations agree very well with the corresponding results obtained with a Coulomb interaction potential, $V_\mathrm{int}(\mathbf{r}_i, \mathbf{r}_j) \sim 1/|\mathbf{r}_i - \mathbf{r}_j|$.
Since $\ell \approx 30~$nm at $\hbar \omega = 1~$meV, this condition is typically well satisfied in the situations considered in the main text.
The resulting lattice structure $\{ \mathbf{r}_1^0, ..., \mathbf{r}_N^0 \}$ depends on the details of the confinement potential.
Two exemplary charge configurations are shown in Fig.~\ref{fig:geometries}.

\subsection{Normal modes}
A two-dimensional lattice with $N$ electrons has $2N$ elementary excitations, the so-called \textit{normal modes} of the crystal.
The normal-mode excitation spectrum of WCs can be calculated from the the system's \textit{elasticity matrix} $\mathcal{K}$.

Starting with Eq.~\eqref{eq:pot-energy-2d-appendix}, the elasticity matrix is obtained from the second-order derivatives of $V_p$ with respect to the spatial coordinates.
In the general case of arbitrary $p \geq 2$ and the interaction potential in Eq.~(2), we find that
\begin{widetext}
\begin{eqnarray}\label{eq:app-2d-hes}
\frac{\partial^2 V_p}{\partial \alpha_m \partial \alpha_n} = \begin{cases}
\mu_p(p-1)p\alpha_m^{p-2} + \xi \left [ \sum_{i\neq m} \frac{(\alpha_i-\alpha_m)^2}{r_0^2} g(|\mathbf{r}_i-\mathbf{r}_m|/r_0) - h(|\mathbf{r}_i-\mathbf{r}_m|/r_0) \right ],
&\text{if $m=n$},\\
-\xi \left [ \frac{(\alpha_n-\alpha_m)^2}{r_0^2} g(|\mathbf{r}_m-\mathbf{r}_n|/r_0) - h(|\mathbf{r}_m-\mathbf{r}_n|/r_0) \right ],
&\text{if $m\neq n$},
\end{cases}
\end{eqnarray}
and
\begin{eqnarray}
\frac{\partial^2 V_p}{\partial \alpha_m \partial \beta_n} = \begin{cases}
\xi \sum_{i\neq m} \frac{(\alpha_i-\alpha_m)(\beta_i-\beta_m)}{r_0^2} g(|\mathbf{r}_i-\mathbf{r}_m|/r_0),
&\text{if $m=n$},\\
- \xi \frac{(\alpha_n-\alpha_m)(\beta_n-\beta_m)}{r_0^2} g(|\mathbf{r}_n-\mathbf{r}_m|/r_0),
&\text{if $m\neq n$},
\end{cases}
\end{eqnarray}
where $\alpha, \beta \in \{ x, y \}$, $\alpha \neq \beta$ and the function $g$ is given by
\begin{equation}
g(x) = H_2(x)+H_{-2}(x) -2H_0(x) - Y_2(x) - Y_{-2}(x) + 2Y_0(x) + \frac{2}{\sqrt{\pi}\Gamma(\frac{3}{2})x} - \frac{x}{2\sqrt{\pi}\Gamma(\frac{5}{2})}.
\end{equation}
\end{widetext}
The eigenmodes of the system are then calculated from the eigenvalues of the elasticity matrix $\mathcal{K}_{mn}^{\alpha \beta} = \partial^2 V_p / (\partial \alpha_m \partial \beta_n)$.

\section{Impurity-induced positional disorder: equidistance measure}\label{app:disorder}
Random dislocations of single electrons from their lattice sites $\mathbf{r}_i^0$ may not only affect the lattice structure of a Wigner crystal, but also the detection scheme and spin couplings discussed in the main text.
For a simple estimate of how severe the impact of impurities on the lattice is, we consider $N_\mathrm{imp}$ randomly distributed Gaussian confinement potentials in addition to the potential in Eq.~\eqref{eq:pot-energy-2d-appendix}, and draw both size and depth of these local confinement potentials from normal distributions.
For our calculations, we assume that they are localized on a nanometer length scale and have a depth of the order of $\sim \mathrm{meV}$.
In a monolayer TMD, such defects could be, e.g., atomistic defects \cite{hu18}.
Starting from Eq.~(1), we take these into account by adding a disorder term,

\begin{align}\label{eq:hamilt-random}
& V(\mathbf{r}_1, ..., \mathbf{r}_N; \{ \mathbf{s}_i \}_{1\leq i\leq N_\mathrm{imp}}) = \frac{m \omega^2}{2}\sum_{i=1}^N \left ( x_i^2 + y_i^2 \right )  \nonumber \\
& + \sum_{i\neq j} V_\mathrm{int} \left ( \mathbf{r}_i, \mathbf{r}_j \right ) + V_\mathrm{rand}(\mathbf{r}_1, ..., \mathbf{r}_N; \{ \mathbf{s}_i \}_{1\leq i\leq N_\mathrm{imp}}),
\end{align}
with
\begin{align}
& V_\mathrm{rand}(\mathbf{r}_1, ..., \mathbf{r}_N; \{ \mathbf{s}_i \}_{1\leq i\leq N_\mathrm{imp}}) = \nonumber \\
& - \sum_{i=1}^N \sum_{j=1}^{N_\mathrm{imp}} \frac{D_\mathrm{j}}{\sqrt{2\pi\sigma_j^2}} \exp \left [ - \frac{(\mathbf{r}_i-\mathbf{s}_j)^T (\mathbf{r}_i-\mathbf{s}_j)}{2\sigma_j^2} \right ],
\end{align}
with random variables $D_j \sim$ meV and $\sigma_j \sim $ nm (where both means and standard deviations are of these orders), where $\{ \mathbf{s}_j \}_{1\leq j\leq N}$ denote the positions of the impurities.
\begin{figure}[b!]
\centering
\begin{overpic}[width=0.98\linewidth]{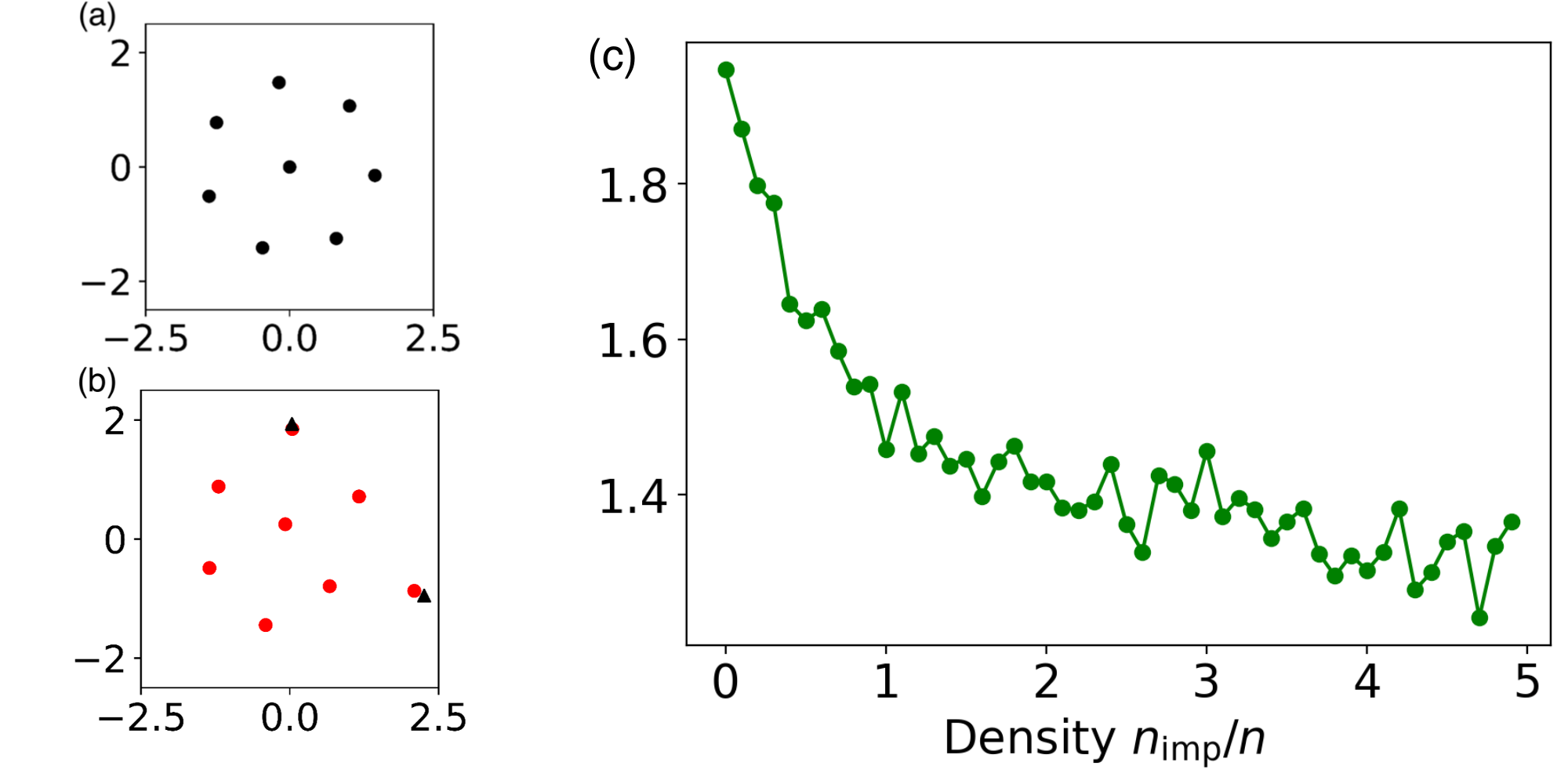}
\put(-2,-2.5){
\begin{tikzpicture}
\path (0,0) node (x) {}
         (0.5,0.5) node[circle,draw,color=white](y){1};
\draw[white] (x) -- (y);
\node (x) at (1.35,0.0) {\textcolor{black}{$x [\ell]$}};
\node (y1) at (0.0,1.25) {\textcolor{black}{$y [\ell]$}};
\node (y2) at (0.0,3.25) {\textcolor{black}{$y [\ell]$}};
\node [rotate=90] (chi) at (3.6,3.0) {$\chi$};
\end{tikzpicture}}
\end{overpic}
\caption{
Impact of disorder-induced potential fluctuations on the lattice structure of a small WC with $N=8$ resident electrons in a harmonic confinement potential.
(\textit{a}) Electron configuration without disorder.
(\textit{b}) Exemplary electron configuration (\textit{red dots}) in the presence of two randomly positioned local confinement potentials (\textit{black triangles}).
(c) Equidistance measure $\chi$ is shown as a function of impurity density $n_\mathrm{imp}/n$ for $N = 10$ electrons.
}
\label{fig:eq-dist}
\end{figure}
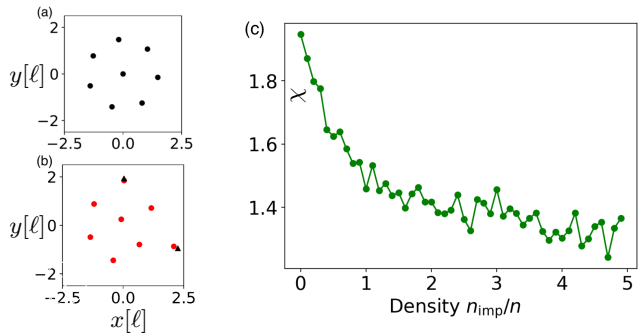
To illustrate how this impurity model affects the lattice site distribution $\mathbf{r}_i^0$ ($i = 1, ..., N$) of a small system, an exemplary numerical result obtained with $N = 8$ is shown in Fig.~\ref{fig:eq-dist}(a).
The same result, but obtained in the presence of two randomly located (in the lattice) local harmonic potentials, is shown in Fig.~\ref{fig:eq-dist}(b).
Averaging over many such instances and calculating the density-density correlations in the WC yields a measure of how much the crystal structure is affected by the presence of disorder.
Similarly, here we look at another measure, $\chi$, which quantifies how \textit{equidistantly} the lattice sites $\mathbf{r}_i^0$ are distributed in the $x$-$y$-plane by summing up the distances between nearest neighbours,
\begin{equation}
    \chi = \frac{2 n}{N} \sum_{i} \underset{j \neq i}{\mathrm{min}} \left | \mathbf{r}_i^0 - \mathbf{r}_j^0 \right |.
\end{equation}
Below we show that $\chi = 2\sqrt{2/\sqrt{3}}$ ($\chi = 1$) for an equidistantly (completely randomly distributed) set of points $\mathbf{r}_i^0$ ($i = 1, ..., N$).
By increasing the number of impurities for a given system size, i.e., increasing the impurity density $n_\mathrm{imp}$ as compared to the electron density $n$, $\chi$ drops from its maximum value very fast, see Fig.~\ref{fig:eq-dist}.
As would be intuitively expected, this underlines that $n_\mathrm{imp} \ll n$ should be fulfilled in any experiment in order to maximize the chances to observe charge ordering in \textit{regular} electron lattices.

We briefly show that $\chi$ is upper-bounded by $\chi_\mathrm{max} = r_\mathrm{m}/r_\infty = 2\sqrt{2}/3^{1/4}\approx 2.15$ \cite{clark54}.
This can be achieved by (i) calculating an upper bound for $r_\mathrm{m} = \sum_{i} \min_{j\not=i}{|\mathbf{r}_i-\mathbf{r}_j|}/N$ and (ii) estimating $r_\infty = 1/(2 \sqrt{n})$ as a function of the average electron density $n$:
(i) In a close-packed lattice with an average nearest-neighbour distance $r_\mathrm{m}$, the unit cell occupies an area in $A_\mathrm{uc} = \sqrt{3} r_\mathrm{m}^2/2$.
The electron density is then given by $n = 2/(\sqrt{3} r_\mathrm{m}^2)$.
(ii) The mean number of lattice sites in a sector of area $A_k=\pi r^2 / k$ is $m = n A_k$. 
The probability of finding $N$ sites in $A_k$ is given by a Poisson distribution $P(N \mathrm{~sites~in~} A_k) = m^N e^{-m} / N!$.
Hence, we obtain the probability that two lattice sites are separated by a distance $|\mathbf{r}_i^0-\mathbf{r}_j^0|$ smaller than a given $r$, $P_<(r):=P(|\mathbf{r}_i^0-\mathbf{r}_j^0|<r)=1-\exp(- n \pi r^2 /k)$.
Therefore, we obtain for the mean of the distance distribution ($k=1$),
\begin{equation}
r_\infty = \int_0^\infty \frac{\mathrm{d} P_<(r)}{\mathrm{d} r} r \mathrm{d} r = \frac{1}{2\sqrt{n}}.
\end{equation}
Combining the findings from (i) and (ii), we obtain an upper bound for $\chi$, $\chi_\mathrm{max} = 2\sqrt{2}/3^{1/4}\approx 2.15$.
Similarly, it can be shown that $\chi = 1$ for a random distribution of lattice sites.

In our numerical calculations, we have seen that the detection scheme is only weakly affected by disorder if the impurity density $n_\mathrm{imp}/n \lesssim 0.1$.
The influence of disorder on cooperative resonances such as the ones discussed in the main text has also been investigated in Ref.~\cite{shahmoon17}.

\begin{figure}[t!]
\begin{overpic}[scale=0.55]{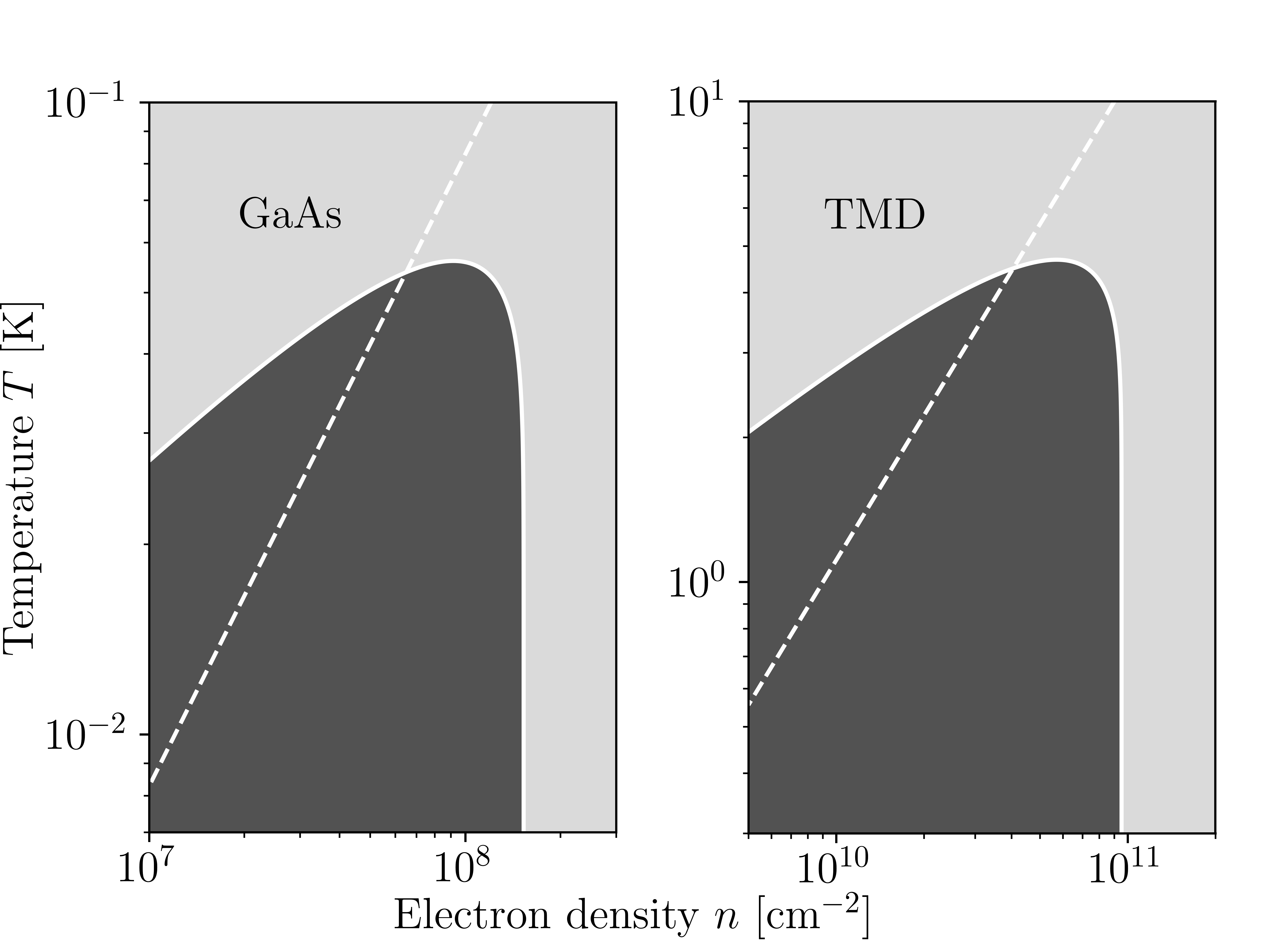}
\end{overpic}
\caption{
Melting curves of \textit{(left)} GaAs and \textit{(right)} monolayer TMD systems according to the Lindemann criterion.
The dark areas indicate the onset of WC electron lattices, obtained for $N = 20$ electrons.
The dashed line indicates $k_\mathrm{B} T = \varepsilon_\mathrm{F}$.
}
\label{fig:phases1}
\end{figure}

\section{Finite temperature effects \label{app:temperature}}
We first provide a simple estimate of the melting temperature $T_\mathrm{m}$ of a WC by employing the \textit{Lindemann criterion}, which has been used extensively in the literature \cite{lindemann10,nozieres58}.
It states that, in a lattice with charge-carrier density $n$, melting occurs if the root-mean square (RMS) displacement of a charge carrier from its lattice site $\mathbf{r}_i^0$ exceeds a certain fraction of the inter-particle distance $a$.
The RMS displacement can be obtained from the thermally occupied vibrational (normal) modes of the system at thermal equilibrium.
Accordingly, the melting temperature $T_\mathrm{m}$ and electron density $n$ can be related.
Although it is only a phenomenological criterion, it provides an efficient tool for estimating the melting temperature of a lattice.
The thereby numerically calculated melting curves, obtained using typical material parameters of GaAs and monolayer TMD systems, respectively, are shown in Fig.~\ref{fig:phases1}.
For the latter, we estimate melting temperatures of the order of $T_\mathrm{m} \sim 5~$K, which is in agreement with previous estimates \cite{zarenia17}.

\begin{figure}[t!]
\begin{overpic}[scale=0.042]{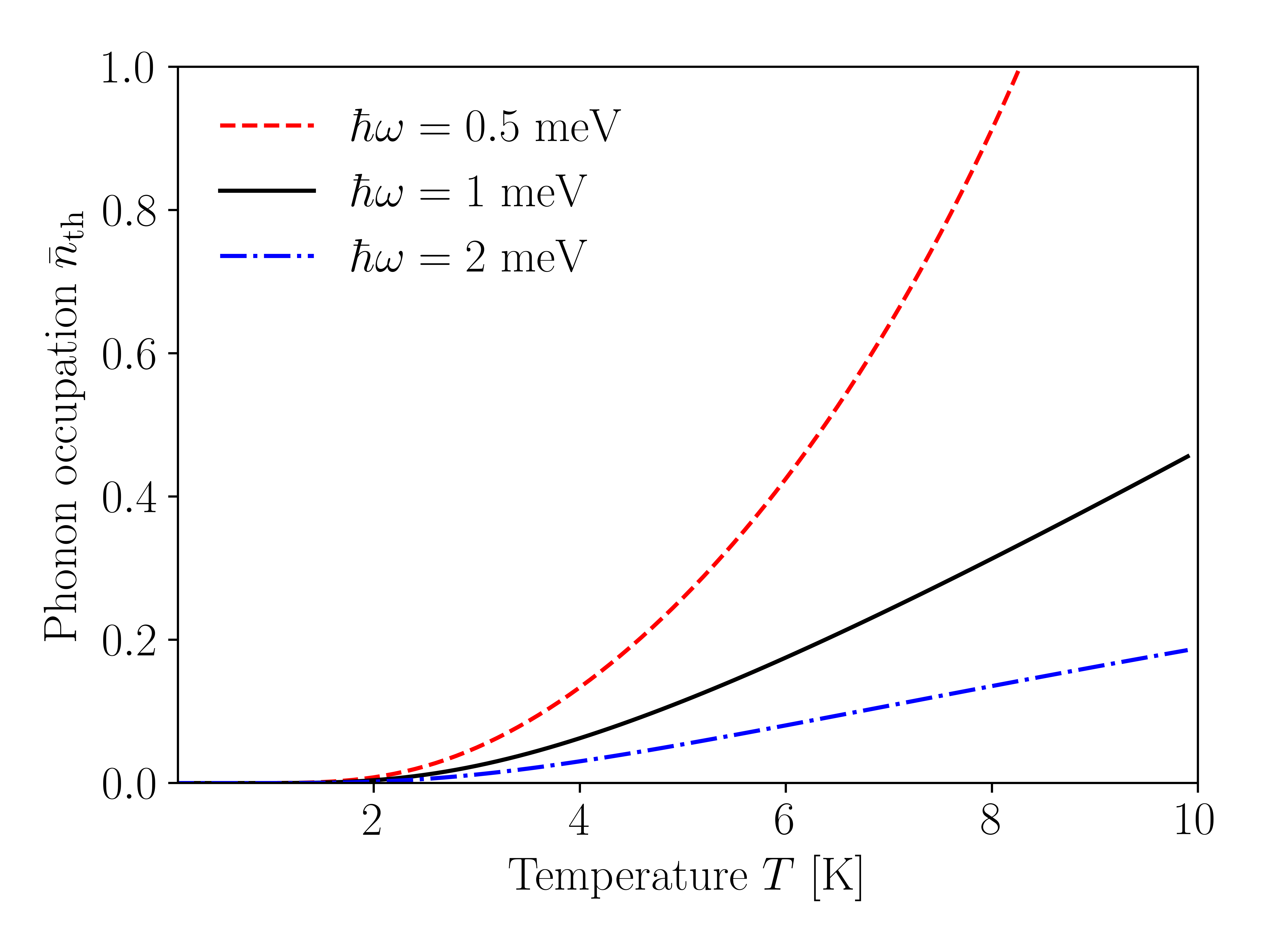}
\end{overpic}
\caption{
Bose-Einstein distribution $\bar n_\mathrm{th} (\Omega_1) $ at COM frequency $\Omega_1 = \omega$ and temperature $T$.
}
\label{fig:bose-einstein}
\end{figure}

Cooling the system into its motional ground state puts more demanding constraints on temperature than considering melting only.
We compare the thermal energy set by $k_\mathrm{B} T$ to the mode frequencies $\Omega_n$ and calculate the thermal mode occupation $\bar n_\mathrm{th} = [\exp(\hbar\Omega_n/(k_\mathrm{B} T) - 1]^{-1}$.
Fig.~\ref{fig:bose-einstein} shows that for the center-of-mass (COM) mode it is $\bar n_\mathrm{th} \ll 1$ at $T \lesssim (1 - 5)~$K and $\hbar\omega\gtrsim 0.5~$meV.

\section{Spin-spin interactions: derivation of coupling constant}\label{app:exchange-coupling}
We estimate the spin-coupling strength $J$ as given by Eq.~(4) in the main text.
For this, we model the interaction potential $V_\mathrm{int}(\mathbf{r}_i, \mathbf{r}_j)$ between two electrons at $\mathbf{r}_i$ and $\mathbf{r}_j$ with a Coulomb potential $\sim 1/|\mathbf{r}_i - \mathbf{r}_j|$.
In the parameter regime considered here, this (i) simplifies the calculation and (ii) yields the same results as obtained with the Keldysh interaction potential from Eq.~(2) to a very good approximation, as confirmed by our numerical calculations.

\subsection{Estimate of spin-coupling constant}
We calculate the spin-exchange interaction between two electrons from the energy difference between the spin-singlet and spin-triplet energies \cite{white},
\begin{equation}\label{eq:app-exchange-constant}
J = \frac{J_{ab}-S^2C}{1-S^4},
\end{equation}
where $J_{ab}$, $C$ and $S$ denote the exchange, Coulomb and overlap
integrals, respectively, which are given by (in atomic units)
\begin{align}\label{eq:definition-JCS-integrals}
J_{ab} & = \int \mathrm{d}^2 \mathbf{r}_1 \int \mathrm{d}^2 \mathbf{r}_2 \Psi_\mathrm{a}(\mathbf{r}_1)^* \Psi_\mathrm{b}(\mathbf{r}_2)^* \frac{1}{|\mathbf{r}_1-\mathbf{r}_2|} \Psi_\mathrm{b}(\mathbf{r}_1) \Psi_\mathrm{a}(\mathbf{r}_2), \nonumber \\ 
S & = \int \mathrm{d}^2 \mathbf{r} \Psi_\mathrm{b}(\mathbf{r}) \Psi_\mathrm{a}(\mathbf{r}), \\
C & = \int \mathrm{d}^2 \mathbf{r}_1 \int \mathrm{d}^2 \mathbf{r}_2 |\Psi_\mathrm{a}(\mathbf{r}_1)|^2 \frac{1}{|\mathbf{r}_1-\mathbf{r}_2|} |\Psi_\mathrm{b}(\mathbf{r}_2)|^2, \nonumber
\end{align}
where $\Psi_{a/b}(\mathbf{r}) = \phi_{a/b} ( \mathbf{r} ) \cdot \chi_{a/b}(\mathbf{r})$ denotes the electronic wave function and the labels $a$ and $b$ refer to the two electrons located at around $\mathbf{r}_{a/b}^0 = ( x_{a/b}^0, y_{a/b}^0 )$.
We model the wave functions with a Gaussian wave packet of width $\textup{w} r_\mathrm{ZPF}$ (see Sec.~S4.2),
\begin{equation}
    \phi_i ( \mathbf{r} ) = \left (\frac{1}{2\pi\textup{w}^2 r_\mathrm{ZPF}^2} \right )^{1/2} \exp \left (-\frac{(x-x_i^0)^2+(y-y_i^0)^2}{4\textup{w}^2 r_\mathrm{ZPF}^2} \right ), \nonumber
\end{equation}
where $i \in \{ a, b \}$, and take into account spin-valley locking by setting the Bloch wave $\chi_i ( \mathbf{r} ) = \exp ( i K x )$ ($\chi_i ( \mathbf{r} ) = \exp ( i K^\prime x )$) if the $i$ electron, $i \in \{ a, b \}$, is in a spin-$\ket{\uparrow}$ (spin-$\ket{\downarrow}$) state.
Next, we evaluate the exchange integral in the spin basis spanned by $\ket{\uparrow \uparrow}, \ket{\uparrow \downarrow}, \ket{\downarrow \uparrow}, \ket{\downarrow \downarrow}$.
With the electrons in different valleys (i.e., opposite spins), by performing some of the integrations analytically, we find for $J_{ab}$ in Eq.~\eqref{eq:definition-JCS-integrals} that
\begin{align}
    J_{ab}^{\mathrm{K}\mathrm{K}^\prime} = & \frac{e^{-\frac{a^2}{4\textup{w}^2 r_\mathrm{ZPF}^2}}}{\pi\textup{w}^2 r_\mathrm{ZPF}^2} \times \nonumber \\
     & \int_{-\infty}^{\infty} \mathrm{d} x e^{- \frac{x^2}{8\textup{w}^2 r_\mathrm{ZPF}^2}} \cos \left ( \frac{8\pi}{3} \frac{x}{a_\mathrm{TMD}} \right ) K_0 \left ( \frac{x^2}{8\textup{w}^2 r_\mathrm{ZPF}^2} \right ), \nonumber
\end{align}
where $a^2 = (x_a^0-x_b^0)^2+(y_a^0-y_b^0)^2$, with a TMD lattice constant $a_\mathrm{TMD} \approx 0.3~$nm for $\mathrm{MoX}_2$ ($\mathrm{X} = \mathrm{S}, \mathrm{Se}$) \cite{rasmussen15} and $|\mathbf{K}-\mathbf{K}^\prime|=8\pi/(3a_\mathrm{TMD})$.
$K_0$ denotes the \textit{modified Bessel function of the second kind}.
Inserting our numerical results for $\textup{w}$, and in particular with $\textup{w} r_\mathrm{ZPF} \gg a_\mathrm{TMD}$, we find numerically that $J_{ab}^{\mathrm{K}\mathrm{K}^\prime}$ evaluates to negligibly small values as compared with $J_{ab}^{\mathrm{K}\mathrm{K}}$, with which we denote the case where the two electrons are in the same valley.
We find that $J_{ab}^{\mathrm{K}\mathrm{K}^\prime}/J_{ab}^{\mathrm{K}\mathrm{K}} \sim a_\mathrm{TMD}/(\mathrm{w}r_\mathrm{ZPF})$ and that typically $J_{ab}^{\mathrm{K}\mathrm{K}^\prime}$ is several orders of magnitude smaller than $J_{ab}^{\mathrm{K}\mathrm{K}}$.

For $J_{ab}^{\mathrm{K}\mathrm{K}}$, we find an analytical expression and insert TMD parameters such that
\begin{equation}\label{eq:Jex-derivation}
J_{ab}^{\mathrm{K}\mathrm{K}} \approx 35.5\mathrm{meV} \ \frac{\sqrt{\hbar \omega \left [ \mathrm{meV} \right ]}}{\textup{w}} \ e^{ - 37.9 \times \frac{\hbar\omega \left [ \mathrm{meV} \right ]}{ n \left [ 10^{10} \mathrm{cm}^{-2} \right ] \textup{w}^2} },
\end{equation}
where we have expressed the electron density as $n = 2/(\sqrt{3}a^2)$ for a triangular lattice.

Similarly, we find for the overlap integral $S$ in Eq.~\eqref{eq:definition-JCS-integrals} that
\begin{equation}\label{eq:S-derivation}
S \approx \exp \left ( - 37.9 \times \frac{\hbar\omega \left [ \mathrm{meV} \right ]}{n \left [ 10^{10} \mathrm{cm}^{-2} \right ]} \right ).
\end{equation}
In the low-density regime considered here, we find $S \ll 1$ such that
$J \approx J_{ab}$ in Eq.~\eqref{eq:app-exchange-constant} to a very good approximation.

Also the Coulomb integral $C$ in Eq.~\eqref{eq:definition-JCS-integrals} can be calculated analytically by employing the convolution and Parseval's theorems.
Defining $f_{a/b}(\mathbf{r}) := |\phi_{a/b}(\mathbf{r})|^2$ and $g(\mathbf{x})=1/|\mathbf{x}|$, we insert TMD parameters and find that
\begin{eqnarray}\label{eq:C-derivation}
C & = & \int \mathrm{d}^2 \mathbf{r}_1 f_\mathrm{A}(\mathbf{r}_1) \left ( f_\mathrm{B} * g \right)(\mathbf{r}_1)
= 2\pi \int \mathrm{d}^2 \mathbf{q} \frac{\widetilde{f_\mathrm{A}}(\mathbf{q}) \widetilde{f_\mathrm{B}}(-\mathbf{q})}{|\mathbf{q}|} \nonumber \\
& \approx & 35.5 \mathrm{meV} \ \frac{\sqrt{\hbar \omega \left [ \mathrm{meV} \right ]}}{\textup{w}} \ e^{ - 18.9 \times \frac{\hbar\omega \left [ \mathrm{meV} \right ]}{n \left [ 10^{10} \mathrm{cm}^{-2} \right ] \textup{w}^2} } \times \nonumber \\
& & I_0\left ( 18.9 \frac{\hbar\omega \left [ \mathrm{meV} \right ]}{n \left [ 10^{10} \mathrm{cm}^{-2} \right ] \textup{w}^2} \right ),
\end{eqnarray}
where $I_0$ is the \textit{modified Bessel function of the first kind}.

Putting our results together, we find that $J^{\mathrm{K}\mathrm{K}^\prime}$ is several orders of magnitude smaller than $J^{\mathrm{K}\mathrm{K}}$ for realistic parameters.
Evaluating the Coulomb interaction Hamiltonian in the spin basis, with these results we obtain the spin model from Eq.~(3) in the main text.
Finally, putting the results from Eqs.~\eqref{eq:Jex-derivation}-\eqref{eq:C-derivation} and Eq.~\eqref{eq:app-exchange-constant} together,
we obtain coupling strengths in the range $\sim (5 - 30)~\mu$eV at densities $n\lesssim n_\mathrm{cr}$, as presented in Fig.~2(a) of the main text.

\subsection{Width of ansatz wavefunction \label{app:width-wavefunction}}
We have considered two approaches to calculate $\textup{w}$, for which we have found good agreement.
$(\textit{i})$ \textit{Mean-field approximation}:
First, we (iteratively, until the result is found to be converged) calculate the effective potential seen by a single electron due to the neighbouring electrons by summing up the Coulomb interaction terms.
From this potential, we calculate the wave function with a Gaussian ansatz, which yields the width of the wave function $\sim \textup{w}$.
$(\textit{ii})$ \textit{Harmonic model}:
Secondly, we consider an expansion of the individual electron displacements in the set of collective displacement modes.
In this way, we relate $\textup{w}$ to the normal modes which we have calculated before,
\begin{equation}
\textup{w}^2 = \frac{1}{N} \sum_{n=1}^{2 N} \frac{1}{\Omega_n}, \ \ \ 
r_\mathrm{ZPF} = \sqrt { \frac{\hbar}{2m\omega} },
\end{equation}
where the mode frequencies $\Omega_n$ are expressed in units of the external confinement $\omega$.
For a confinement $\hbar\omega = 3~$meV, we obtain $r_\mathrm{ZPF} \approx 5~$nm.

\section{Optical readout: numerical and analytical treatment}\label{app:optical-readout}

Here we first briefly summarize how we solve the scattering problem of light incident on a finite Wigner crystal, and then continue with an analytical treatment of the scattering problem for an infinite lattice.
The latter provides us with more physical insight into the problem and is useful for optimizing the beam parameters in order to maximize the transmission or reflection contrast of the readout scheme.

\subsection{Finite arrays}

The principle behind the optical readout scheme discussed in the main text is based on a cooperative resonance effect as described in detail in Refs.~\cite{shahmoon17,bettles16}.
As depicted in Fig.~1, we consider a Gaussian beam $\mathbf{E}_\mathrm{in}(x^\prime, y^\prime, z^\prime)$ incident on the $xy$ plane with a tilt angle $\theta$ and azimuthal angle $\phi$, where
\begin{align}\label{eq:gaussian-beam}
    \mathbf{E}_\mathrm{in}(x, y, z) = & E_0 \mathbf{e}_\mathrm{pol} \frac{w_0}{w(z)} \exp \left (-\frac{x^2+y^2}{w(z)^2} \right ) \times \\
    & \exp \left (-i \left [ kz + k\frac{x^2+y^2}{R(z)} - \varphi(z) \right ] \right ), \nonumber
\end{align}
that is scattered from a lattice of dipoles.
Here we have introduced the coordinates
\begin{equation}
  \begin{pmatrix}
  x^\prime\\
  y^\prime\\
  z^\prime
  \end{pmatrix} = 
  \begin{pmatrix}
    x\cos\theta\cos\phi-y\cos\theta\sin\phi-z\sin\theta\\
    x\sin\phi + y\cos\phi\\
    x\sin\theta\cos\phi-y\sin\theta\sin\phi+z\cos\theta
  \end{pmatrix}.
\end{equation}
In Eq.~\eqref{eq:gaussian-beam}, $E_0$ denotes the the beam amplitude, $w_0$ and $w(z)=w_0 \sqrt{1 + (z/z_\mathrm{R})^2}$ are beam waist and radius at $z$, respectively, $z_\mathrm{R} = \pi w_0^2 / \lambda$ is the \textit{Rayleigh length} and $\varphi = \arctan z/z_\mathrm{R}$ refers to the \textit{Gouy phase} of the laser beam \cite{svelto}.
$\mathbf{e}_\mathrm{pol}$ encodes the polarization of the beam.
For the results presented in Fig.~3 in the main text we consider elliptically polarized light with
\begin{equation}\label{eq:ellip-pol}
\mathbf{e}_\mathrm{pol}(\theta, \phi) = - \frac{1}{\sqrt{1+\cos^2\theta}}
\begin{pmatrix}
\cos^2\theta \cos \phi + i \sin \phi\\
\cos^2\theta \sin \phi - i \cos \phi\\
\sin\theta\cos\theta
\end{pmatrix}.
\end{equation}
At small detunings $\Delta_0$ from the transition frequency $\omega_0$, $|\Delta_0| \ll \omega_0$, each lattice site is modeled as a dipole with polarizability
\begin{equation}\label{eq:polarizability-appendix}
\alpha(\Delta_0) = - \frac{3}{8\pi^2} \varepsilon_0 \lambda^3 \frac{\gamma_\mathrm{r}}{\Delta_0 + i (\gamma_\mathrm{r} + \gamma_\mathrm{nr})/2},
\end{equation}
with the radiative (nonradiative) linewidth $\gamma_\mathrm{r}$ ($\gamma_\mathrm{nr}$).
In general, the radiative linewidth $\gamma_\mathrm{r}$ can be enhanced by the presence of a medium \cite{thranhardt02}, especially for high refractive-index materials like TMDs \cite{liu14}.
At low temperatures as considered here, hexagonal boron nitride (hBN) encapsulated TMD monolayers feature optical transitions with a radiative linewidth $\hbar \gamma_\mathrm{0} \sim \mathrm{meV}$ \cite{fang19}.
In our calculations, we assume that the excitons are localized on a length scale much smaller than the wavelength, i.e.~$a_\mathrm{B} \ll \lambda$.
Those spatially localized quantum emitters show much narrower linewidths $\sim 100~\mu$eV \cite{srivastava15,he15,koperski15,chakraborty15}.
Using Fermi's golden rule, the increased radiative lifetime of such localized excitons can be calculated, yielding a significantly enhanced emission time as compared to free excitons \cite{ayari18}.
We estimate the radiative linewidth of a localized exciton to be of the order of $\hbar \gamma_\mathrm{r} \approx 4\pi/3 (a_\mathrm{B}/\lambda)^2 \gamma_0 \approx 10^{-5}~\gamma_0 \approx 10^{-2}~\mu$eV.
In the results presented in the main text, we have considered $\gamma_\mathrm{nr} = 0$.

\begin{figure}[b!]
  \centering
  \includegraphics[width=0.99\linewidth]{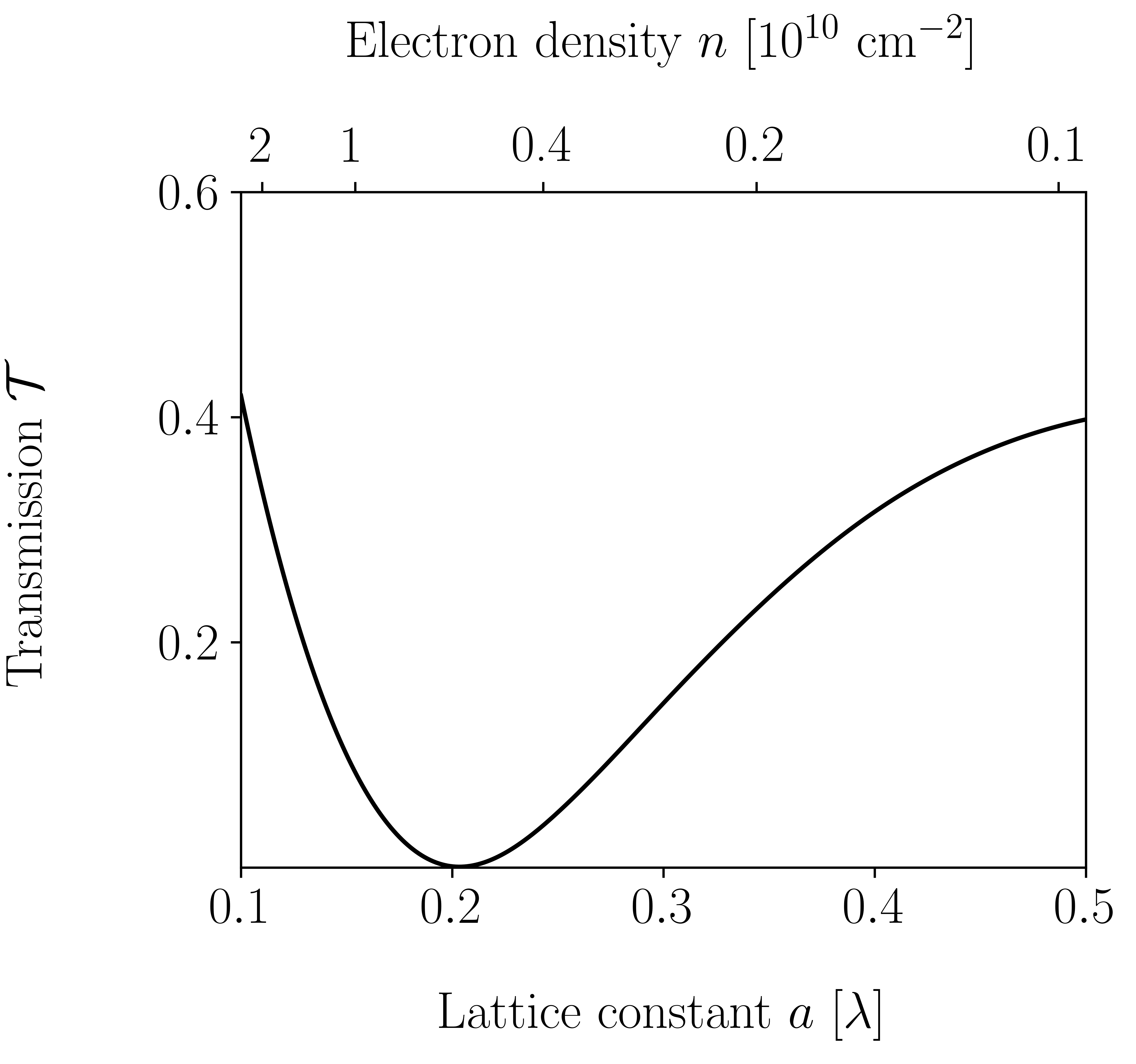}
  \caption{Transmission at normal incidence ($\theta = 0$) for a square lattice. Other numerical parameters as in Fig.~3 in the main text.}
  \label{fig:theta-zero}
\end{figure}

Given the Gaussian input field from Eq.~\eqref{eq:gaussian-beam}, we solve the Lippmann-Schwinger equation (4), with the Green's function \cite{novotny06}
\resizebox{.485\textwidth}{!}{
  \begin{minipage}{0.5\textwidth}
\begin{align}\label{eq:greens}
G_{\alpha\beta}(k, \mathbf{r}, \mathbf{r}_n^0) = & \frac{\exp \left ( i k \left | \mathbf{r} - \mathbf{r}_n^0 \right | \right )}{4\pi\left | \mathbf{r} - \mathbf{r}_n^0 \right |} \times \Bigg [ \left ( 1 + \frac{ik\left | \mathbf{r} - \mathbf{r}_n^0 \right |}{k^2\left | \mathbf{r} - \mathbf{r}_n^0 \right |^2} \right ) \delta_{\alpha\beta} \ + \nonumber \\
& \left ( \frac{3 - 3ik\left | \mathbf{r} - \mathbf{r}_n^0 \right |}{k^2\left | \mathbf{r} - \mathbf{r}_n^0 \right |^2} - 1 \right ) \frac{\left ( \mathbf{r} - \mathbf{r}_n^0 \right )_\alpha \left ( \mathbf{r} - \mathbf{r}_n^0 \right )_\beta}{\left | \mathbf{r} - \mathbf{r}_n^0 \right |^2} \Bigg ], \nonumber
\end{align}
\end{minipage}
}
\newline\newline
\noindent with $\alpha, \beta \in \{ x, y, z \}$.
We solve Eq.~(4) self-consistently for various angles of incidence $\theta$ and $\phi$, beam profiles, detunings, and electron lattices.
At normal incidence, i.e.~$\theta = 0$, the resulting transmission and reflection signals depend on the lattice constant [see Fig.~\ref{fig:theta-zero}] but clearly not on $\phi$.
For $0 < \theta < \pi/2$, the transmission and reflection contrasts can be of the order of a few percent.
An analytical derivation of the maximum contrast for an infinite lattice, depending on the angle of incidence $\theta$ and detuning $\Delta_0$, is presented in Sec.~\ref{ssec:infinite-arrays}.

\subsection{Faraday rotation}

In the main text we investigate the Faraday rotation angle according to Eq.~(6).
For the results in Fig.~4, we consider an incoming beam at normal incidence ($\theta = 0$) with
\begin{equation}
\mathbf{e}_\mathrm{pol} = \begin{pmatrix}1\\ 0\\ 0\end{pmatrix}.
\end{equation}
We consider $N \equiv N_\uparrow + N_\downarrow$ dipoles which are located at lattice sites $\mathbf{r}_i^0$ with the spins assigned randomly to these lattice lattices for fixed $N_\uparrow$ and $N_\downarrow$.
Next we average over sufficiently many ($\sim 10^4$) instances of such configurations to calculate the Faraday rotation.

\begin{figure}[t!]
  \centering
  \includegraphics[width=1.0\linewidth]{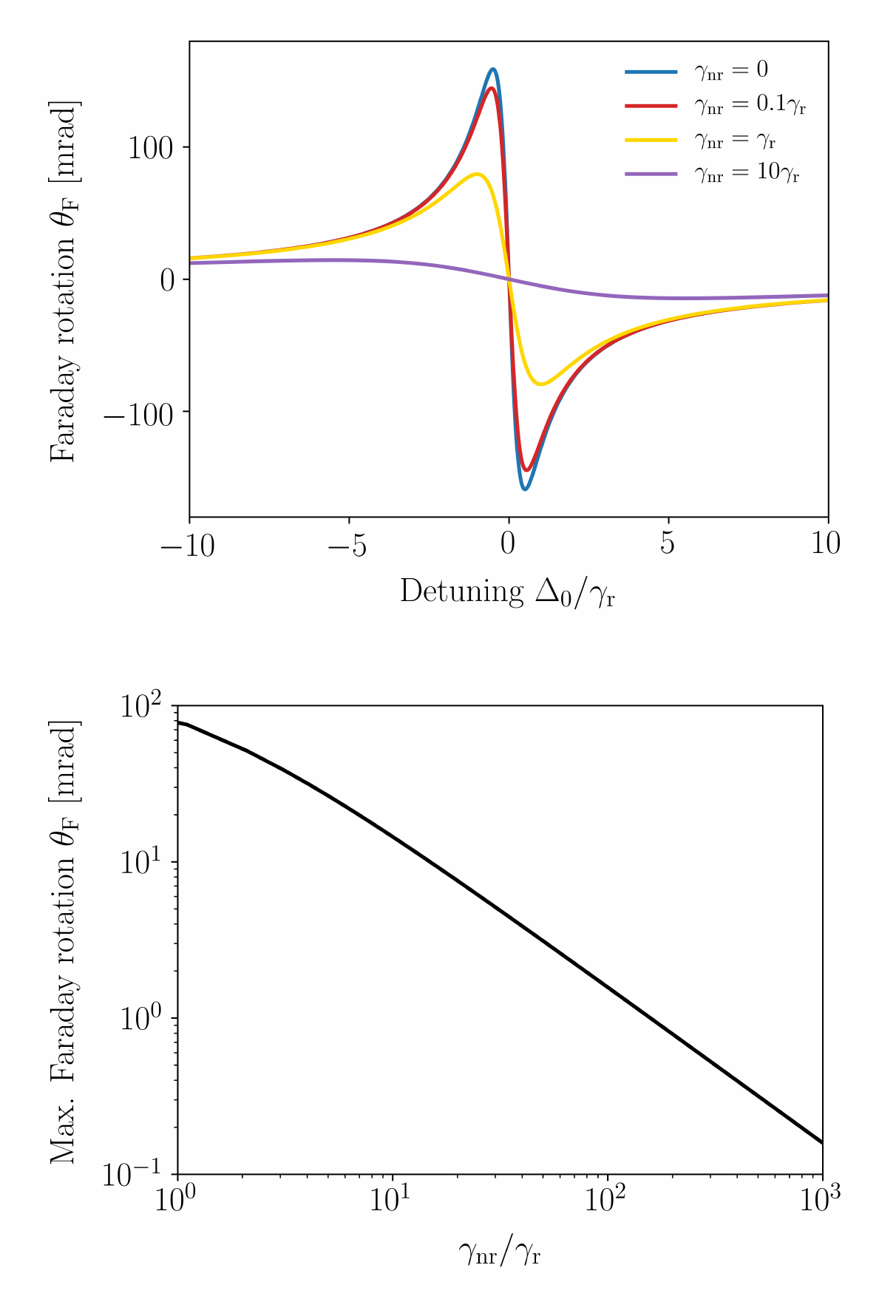}
  \caption{Faraday rotation for different $\gamma_\mathrm{nr}$ and the same numerical parameters as in Fig.~3 at $N_\uparrow = 15$, $N_\downarrow = 5$. Also shown is the maximum Faraday signal as a function of $\gamma_\mathrm{nr}/\gamma_\mathrm{r}$.}
  \label{fig:gamma_nr}
\end{figure}

In Fig.~4 we show results for $\gamma_\mathrm{nr} = 0$.
For $\gamma_\mathrm{nr} > 0$, the maximum Faraday rotation decreases and shifts towards more highly detuned frequencies, cf.~Fig.~\ref{fig:gamma_nr}.

\subsection{Infinite arrays \label{ssec:infinite-arrays}}

Here we consider light scattering off an (infinite) two-dimensional lattice of dipoles.
If the transition dipole is parallel to the unit vector $\vec{\hat e}$, the electric field at position $\vec{r}$ satisfies the equation
\begin{equation}
  \vec{E}(\vec{r}) = \vec{E}_\mathrm{in}(\vec{r}) + \alpha(\Delta_0) \frac{k_0^2}{\varepsilon_0} \sum_n G(\vec{r}, \vec{r}_n) \vec{\hat e} \vec{\hat e}^\dagger \vec{E}(\vec{r}_n),
\end{equation}
where $k_0$ denotes the wavenumber of the transition.
This equation can be readily solved using a Fourier transform, assuming that the medium surrounding the lattice is translationally invariant in the plane of the lattice.
One obtains
\begin{align}\label{eq:solution}
  \vec{E}(\vec{k}, z) = & \vec{E}_\mathrm{in}(\vec{k}, z) + \alpha(\Delta_0) \frac{k_0^2}{\varepsilon_0 A} G(\vec{k}, z) \vec{\hat e} \vec{\hat e}^\dagger \times \nonumber \\
  & \left[ \mathbb{I} - \alpha(\Delta_0) \frac{k_0^2}{\varepsilon_0 A} \tilde G(\vec{k}) \vec{\hat e} \vec{\hat e}^\dagger \right]^{-1} \sum_\vec{B} \vec{E}_\mathrm{in}(\vec{k} + \vec{B}, 0)
\end{align}
where $A$ is the area of the unit cell and
\begin{equation}
  \tilde G(\vec{k}) = \sum_\vec{B} G(\vec{k} + \vec{B}, 0),
\end{equation}
where the sum runs over all reciprocal lattice vectors, denoted by $\vec{B}$.

For an incident plane wave with momentum $\vec{k}$, only a single term contributes to the sum in Eq.~\eqref{eq:solution}. The plane wave will be Bragg scattered to momenta $\vec{k} + \vec{B}$. However, for sufficiently small lattice constants, $|\vec{k} + \vec{B}| > k_0$ for any $\vec{B} \neq 0$, such that all nonzero scattering orders are evanescent. In this case, the far field is completely described by
\begin{align}\label{eq:far_field}
  \vec{E}(\vec{k}, z) = \vec{E}_\mathrm{in}(\vec{k}, z) - & \frac{3 \pi \gamma_\mathrm{r} / k_0 A}{\Delta_0 + i \gamma_\mathrm{nr}/2 + 3 \pi \gamma_\mathrm{r} \vec{\hat e}^\dagger \tilde G(\vec{k}) \vec{\hat e}/k_0 A} \times \nonumber \\
  & G(\vec{k}, z) \vec{\hat e} \vec{\hat e}^\dagger \vec{E}_\mathrm{in}(\vec{k} , 0).
\end{align}
We were able to turn the matrix inversion into a simple division by using the fact that $\vec{\hat e} \vec{\hat e}^\dagger$ is a projector.
It is straightforward to show that the condition $|\vec{k} + \vec{B}| > k_0$ is equivalent to $|\vec{B}| > 4 \pi / \lambda$.
For a square lattice, one obtains $a < \lambda/2$, while for a triangular lattice $a < \lambda / \sqrt{3}$.

To simplify Eq.~\eqref{eq:far_field} further, we consider the special case that the array is placed in free space.
The free space Green's function is given by
\begin{equation}
  G(\vec{k}, z) = \frac{i}{2 k_z} e^{i k_z |z|} P_\pm(\vec{k}),
\end{equation}
where
\begin{equation}
  k_z = \sqrt{k_0^2 - |\vec{k}|^2}
\end{equation}
and $P_\pm(\vec{k})$ denotes the projector onto transverse polarizations for waves propagating up ($+$, $z > 0$) or down ($-$, $z < 0$). Explicitly, the $P_\pm(\vec{k})$ projects onto the two-dimensional space spanned by
\begin{equation}
  \vec{\hat s}(\phi) = 
  \begin{pmatrix}
    -\sin \phi\\
    \cos \phi\\
    0
  \end{pmatrix}, \qquad
  \vec{\hat p}_\pm(\theta, \phi) = 
  \begin{pmatrix}
    \pm \cos \theta \cos \phi\\
    \pm \cos \theta \sin \phi\\
    - \sin \theta
  \end{pmatrix},
\end{equation}
where we defined the angles $\theta$ and $\phi$ according to
\begin{equation}
  k_x = k_0 \sin \theta \cos \phi, \quad
  k_y = k_0 \sin \theta \sin \phi, \quad
  k_z = k_0 \cos \theta.
\end{equation}
We note that $k_z$ is always taken to have a positive real ($|\vec{k}| < k_0$) or imaginary ($|\vec{k}| > k_0$) part.
When $|\vec{k}| < k_0$, all angles are real, and the vector $(k_x, k_y, -k_z)$ is simply the wavevector of the incident wave.
We also point out that the Green's function is discontinuous at $z = 0$. Right at $z = 0$, one should take \footnote{We further neglect an unimportant $\delta$-function contribution.}
\begin{equation}
  G(\vec{k}, 0) = \frac{i}{4 k_z} e^{i k_z |z|}\left [ P_+(\vec{k}) + P_-(\vec{k}) \right].
\end{equation}

We focus on a circularly polarized transition, that is,
\begin{equation}
  \vec{\hat e} = \frac{1}{\sqrt{2}}
  \begin{pmatrix}
    1\\
    i\\
    0
  \end{pmatrix}.
\end{equation}
When there is no Bragg scattering, it is easy to see that $\im \, \tilde G(\vec{k}) = \im \, G(\vec{k}, 0)$ such that
\begin{equation}
  \vec{\hat e}^\dagger \im \, \tilde G(\vec{k}) \vec{\hat e} = \frac{i}{4 k_0} \frac{1 + \cos^2 \theta}{\cos \theta}.
\end{equation}
A straightforward calculation further yields
\begin{equation}
  P_+(\vec{k}) \vec{\hat e} \vec{\hat e}^\dagger P_-(\vec{k}) = \frac{1}{2} (1 + \cos^2 \theta) \vec{\hat v}_+(\theta, \phi) \vec{\hat v}_-(\theta, \phi)^\dagger,
\end{equation}
where
\begin{equation}
  \vec{\hat v}_\pm(\theta, \phi) = \frac{1}{\sqrt{1 + \cos^2 \theta}} \left[ i \vec{\hat s}(\phi) \pm \cos \theta \vec{\hat p}_\pm(\theta, \phi) \right].
\end{equation}
Since $\vec{E}_\mathrm{in}(\vec{k}) = P_-(\vec{k}) \vec{E}_\mathrm{in}(\vec{k})$, we thus obtain
\begin{align}
  \label{eq:final}
  \vec{E}(\vec{k}, z) &= \\
  &\Bigg[ e^{- i k_z z} - e^{i k_z |z|} \frac{i \Gamma(\theta)/2}{\Delta_0 + \tilde \Delta(\theta, \phi) + i \gamma_\mathrm{nr}/2 + i \Gamma(\theta)/2} \times \nonumber \\
  & \ \vec{\hat v}_\pm(\theta, \phi) \vec{\hat v}_-(\theta, \phi)^\dagger \Bigg] \vec{E}_\mathrm{in}(\vec{k} , 0),\nonumber
\end{align}
where
\begin{equation}
  \Gamma(\theta) = \frac{3 \pi \gamma_\mathrm{r}}{2 k_0^2 A} \frac{1 + \cos^2 \theta}{\cos \theta}
\end{equation}
and
\begin{equation}
  \tilde \Delta(\theta, \phi) = \frac{3 \pi \gamma_\mathrm{r}}{k_0 A} \vec{\hat e}^\dagger \re \, \tilde G(\vec{k}) \vec{\hat e}.
\end{equation}

Eq.~\eqref{eq:final} has a simple physical interpretation.
The light probes a collective resonance with energy $\tilde \Delta$ and radiative linewidth $\Gamma$.
The vectors $\vec{\hat v}_\pm$ correspond to projections of the transverse polarizations onto the transition dipole.
The response of the lattice is maximized when $\vec{E}_\mathrm{in} \propto \vec{\hat v}_-$, which corresponds to an elliptic polarization whose projection onto the $xy$ plane is circular.
The expression allows us to immediately read off the reflection and transmission coefficients:
\begin{equation}
  r = - \frac{i \Gamma(\theta) / 2}{\Delta_0 + \tilde \Delta(\theta, \phi)+ i \gamma_\mathrm{nr} / 2 + i \Gamma(\theta) / 2} \vec{\hat v}_+ (\theta, \phi) \vec{\hat v}_-(\theta, \phi)^\dagger,
\end{equation}
\begin{equation}
  t = P_- - \frac{i \Gamma(\theta) / 2}{\Delta_0 + \tilde \Delta(\theta, \phi)+ i \gamma_\mathrm{nr} / 2 + i \Gamma(\theta) / 2} \vec{\hat v}_- (\theta, \phi) \vec{\hat v}_-(\theta, \phi)^\dagger.
\end{equation}
Both $r$ and $t$ should be thought of as $2 \times 2$ matrices acting on the subspaces of transverse polarizations. For a fixed incident polarization $\vec{\hat e}_\mathrm{in}$, we may further compute the intensity reflection and transmission cofficients. They are given by
\begin{equation}
  R = \frac{\Gamma(\theta)^2 / 4}{ [ \Delta_0 + \tilde \Delta(\theta, \phi)]^2 + [\gamma_\mathrm{nr} + \Gamma(\theta)]^2 / 4} \left| \vec{\hat v}_-(\theta, \phi)^\dagger \vec{\hat e}_\mathrm{in} \right|^2,
\end{equation}
\begin{equation}
  T = 1 - \frac{\Gamma(\theta) [ \Gamma(\theta) + 2 \gamma_\mathrm{nr} ] / 4}{ [ \Delta_0 + \tilde \Delta(\theta, \phi)]^2 + [\gamma_\mathrm{nr} + \Gamma(\theta)]^2 / 4} \left| \vec{\hat v}_-(\theta, \phi)^\dagger \vec{\hat e}_\mathrm{in} \right|^2.
\end{equation}
The intensity coefficients satisify $R + T = 1$ when $\gamma_\mathrm{nr} = 0$ as required.

\begin{figure}[b!]
  \centering
  \includegraphics{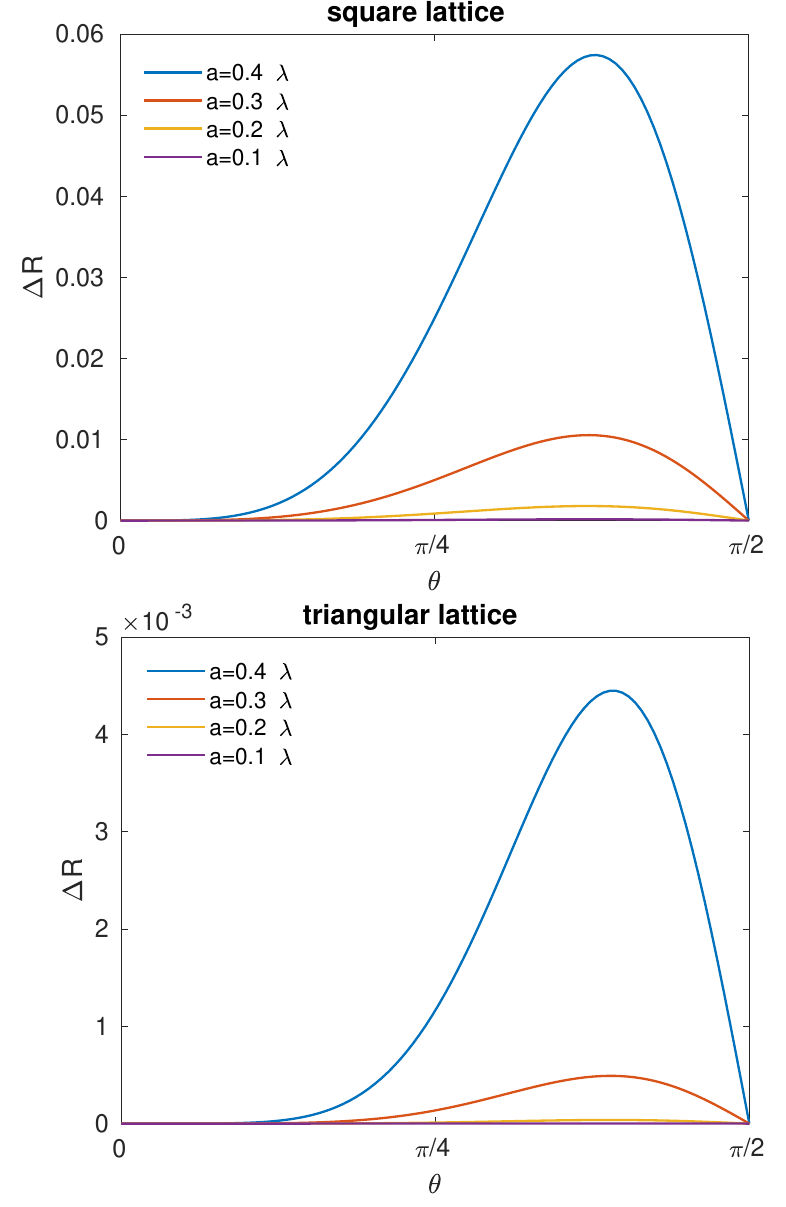}
  \caption{Reflection contrast according to Eq.~\eqref{eq:approx} for a square and triangular lattice and various lattice constants $a$.}
  \label{fig:theta}
\end{figure}
In practice, we would like to infer the rotational symmetry of the lattice via the dependence of $\tilde \Delta$ on $\phi$. Choosing the optimal polarization $\vec{\hat e}_\mathrm{in} = \vec{\hat v}_-(\theta, \phi)$, the maximum contrast in reflection for a fixed value of $\theta$ is given by
\begin{equation}
  \Delta R = \frac{\Gamma^2}{4} \left[ \frac{1}{(\Delta_0 + \tilde \Delta_\mathrm{min})^2 + \Gamma^2 / 4} - \frac{1}{(\Delta_0 + \tilde \Delta_\mathrm{max})^2 + \Gamma^2 / 4} \right]
\end{equation}
where $\tilde \Delta_\mathrm{min} = \min_\phi \tilde \Delta(\theta, \phi)$ and similarly for $\tilde \Delta_\mathrm{max}$. For simplicity we set $\gamma_\mathrm{nr} = 0$, which implies that the contrast in transmission is equal to the contrast in reflection. We are free to choose $\Delta_0$ to maximize the contrast. Writing $\Delta_0 = -(\tilde \Delta_\mathrm{min} + \tilde \Delta_\mathrm{max})/2 + \delta$, the contrast can be expressed as
\begin{equation}
  \Delta R = \frac{\delta \bar \Delta / \Gamma^2}{(\delta^2/\Gamma^2 - \bar \Delta^2/\Gamma^2 +1/4)^2 + \bar \Delta^2 / \Gamma^2},
\end{equation}
where $\bar \Delta = (\tilde \Delta_\mathrm{max} - \tilde \Delta_\mathrm{min})/2$. In the limit $\bar \Delta \ll \Gamma$, the expression simplifies to
\begin{equation}
  \Delta R \approx \frac{\delta \bar \Delta / \Gamma^2}{(\delta^2 / \Gamma^2 + 1/4)^2}
\end{equation}
It is easy to show that the contrast is maximized by choosing
\begin{equation}
  \delta = \frac{1}{2 \sqrt{3}} \Gamma,
\end{equation}
yielding
\begin{equation}
  \Delta R \approx \frac{3 \sqrt{3}}{2} \frac{\bar \Delta}{\Gamma}.
  \label{eq:approx}
\end{equation}
The value of $\bar \Delta$ can be computed numerically. The results for a square and triangular lattice are plotted below.

\begin{figure}[t!]
  \centering
  \includegraphics[height=6cm]{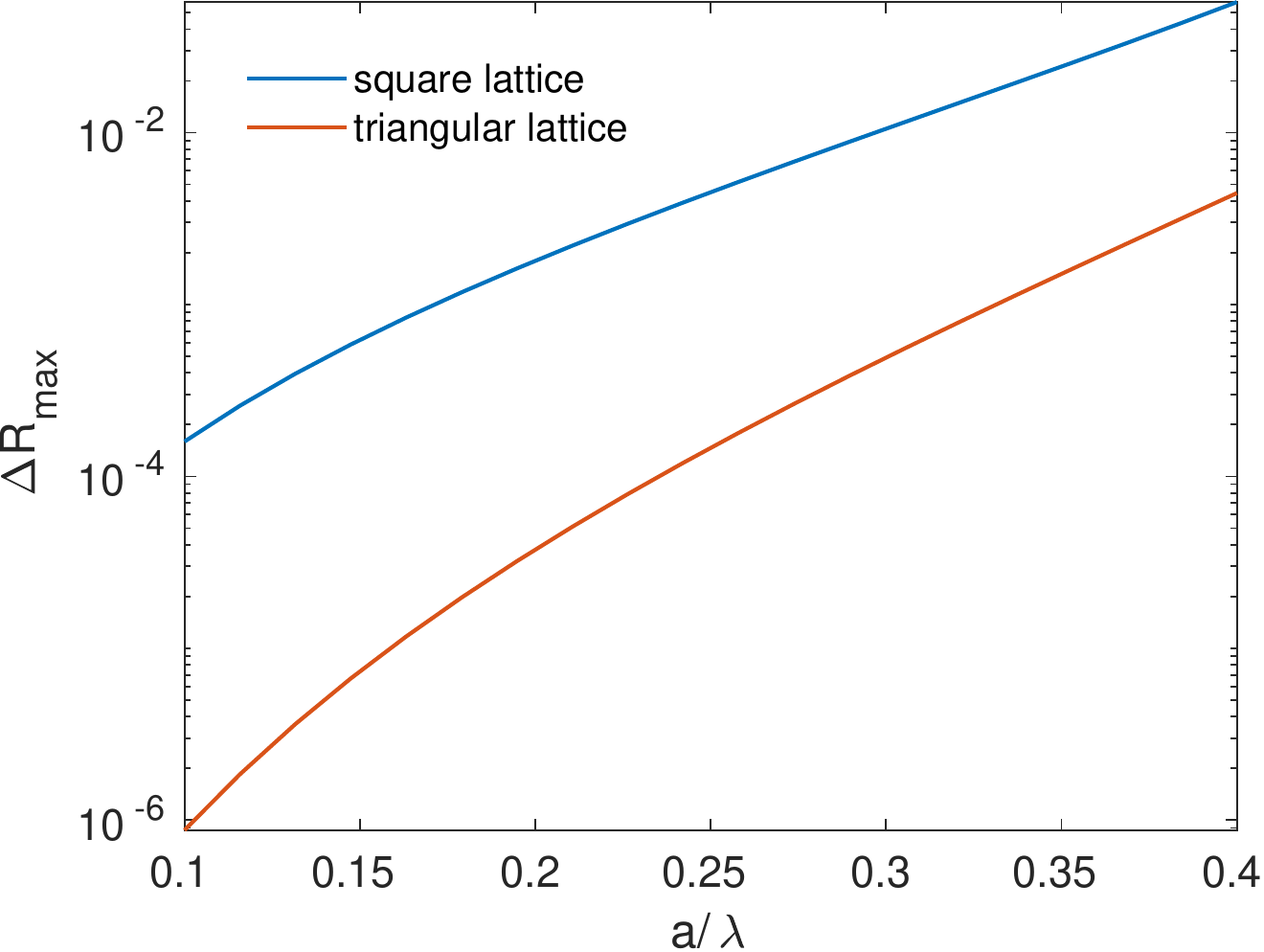}
  \caption{Maximum value of $\Delta R$ as a function of lattice constant.}
  \label{fig:r}
\end{figure}

As a final remark, we mention that by measuring the transmission coefficient for a component of the electric field that is neither parallel nor perpendicular to the incident field, it is possible to observe dispersive (asymmetric) line shapes. Such features could potentially enhance the sensitivity.


\end{document}